\documentclass[aps,pre,twocolumn,showpacs,amssymb,superscriptaddress,floatfix]{revtex4}

\usepackage[dvips]{epsfig,graphicx}
\usepackage{amsmath,amssymb,amsfonts,wasysym,color}

\def\be{\begin{equation}}
\def\ee{\end{equation}}
\def\la{\langle}
\def\ra{\rangle}

\def\Cas{\mathrm{C}}
\def\bulk{\mathrm{bulk}}

\def\ex{\mathrm{ex}}
\def\kB{k_{\mathrm{B}}}
\def\omegaeff{\omega_{{\rm eff}}}

\def\C{{\mathcal C}}

\def\rme{{\rm e}}
\def\rmd{{\rm d}}
\newcommand\reff[1]{(\ref{#1})}
\def\bea{\begin{eqnarray}}
\def\eea{\end{eqnarray}}

\def\sf{\vartheta}
\def\sfh{\hat\vartheta}
\def\sfb{\theta}

\def\sfhb{\hat\theta}
\begin{document}
\title{
Universal scaling functions  of critical Casimir forces obtained by Monte
Carlo simulations %
}
\author{O.~Vasilyev}
\affiliation{Max-Planck-Institut f{\"u}r Metallforschung, Heisenbergstr.~3, D-70569 Stuttgart, Germany}
\affiliation{Institut f{\"u}r Theoretische und Angewandte Physik,
Universit{\"a}t Stuttgart, Pfaffenwaldring 57, D-70569 Stuttgart, Germany}
\author{A.~Gambassi}
\affiliation{Max-Planck-Institut f{\"u}r Metallforschung, Heisenbergstr.~3, D-70569 Stuttgart, Germany}
\affiliation{Institut f{\"u}r Theoretische und Angewandte Physik,
Universit{\"a}t Stuttgart, Pfaffenwaldring 57, D-70569 Stuttgart, Germany}
\author{A.~Macio\l ek}
\affiliation{Max-Planck-Institut f{\"u}r Metallforschung, Heisenbergstr.~3, D-70569 Stuttgart, Germany}
\affiliation{Institut f{\"u}r Theoretische und Angewandte Physik,
Universit{\"a}t Stuttgart, Pfaffenwaldring 57, D-70569 Stuttgart, Germany}
\affiliation{Institute of Physical Chemistry,
             Polish Academy of Sciences, Kasprzaka 44/52,
            PL-01-224 Warsaw, Poland}
\author{S.~Dietrich}
\affiliation{Max-Planck-Institut f{\"u}r Metallforschung, Heisenbergstr.~3, D-70569 Stuttgart, Germany}
\affiliation{Institut f{\"u}r Theoretische und Angewandte Physik,
Universit{\"a}t Stuttgart, Pfaffenwaldring 57, D-70569 Stuttgart, Germany}
\date{\today}

\begin{abstract}
Effective Casimir forces induced by thermal fluctuations
in the vicinity of bulk critical points are studied  by means of
Monte Carlo simulations in three-dimensional systems for film geometries 
and within the experimentally
relevant Ising and XY universality classes.  
Several surface universality classes of the confining surfaces
are considered, some of  which are relevant for recent experiments.
A novel approach introduced  previously 
[EPL {\bf 80}, 60009 (2007)], 
 based inter alia on an integration scheme of free energy
differences, is utilized to compute the universal 
scaling functions of the critical Casimir 
forces in the critical range of temperatures above and below the bulk 
critical temperature. The resulting predictions  are compared 
with corresponding experimental data for wetting films of fluids 
 and with available theoretical results.
\end{abstract}


\maketitle

\section{Introduction}
\label{sec:intr}

The confinement of a fluctuating medium generates 
effective forces acting on the corresponding surfaces.
Close to the critical point of a continuous phase transition the
relevant fluctuating degree of freedom is the order
parameter of the phase transition. The effective 
force resulting from the confinement
of such critical fluctuations is known as the {\it critical Casimir force}
$f_\Cas$. 
This force has a universal character
in the sense that it is largely
independent of the microscopic details of the systems and of the confining
surfaces but depends only on some of their gross features (which
characterize the corresponding bulk and surface 
universality classes), as it is typically the case for bulk and surface 
critical phenomena. Such forces were first discussed by
 Fisher and de Gennes \cite{FdG} on the basis of  finite-size
 scaling ~\cite{FSS} for a fluid  system confined by two parallel walls.

After early qualitative observations~\cite{footnote1,bey}
the first {\it quantitative}
experimental evidence for such a force
was provided by the study of
wetting layers of $^4$He~\cite{garcia}, 
where $f_\Cas$ originates from the confined critical
fluctuations associated with the superfluid transition in the fluid film;
$f_\Cas$ adds to the omnipresent background dispersion forces which together
determine the equilibrium thickness $L$ of the
wetting layers~\cite{garcia}. 
The dependence of  $L$ on temperature $T$ provides an indirect
measurement of $f_\Cas$; varying the undersaturation allows one to tune $L$
and thus to probe the scaling properties of $f_\Cas$ as function of $T$ and
$L$~\cite{krech:92,KD-92}. 
Later on, wetting layers of classical~\cite{pershan,rafai} 
and quantum binary liquid mixtures~\cite{chan:mix} have been
 studied and in two
cases it has been possible to determine quantitatively the critical Casimir
force near a critical~\cite{pershan} and a tricritical~\cite{chan:mix} point.
Only recently, however, the 
existence of the critical Casimir effect  has been demonstrated by 
a {\it direct}  measurement of the  femto-Newton force between a planar wall 
and a colloidal particle  immersed in a  near-critical binary liquid
mixture~\cite{Hertlein}.

The universality of the Casimir force $f_\Cas$ allows one to investigate
its temperature dependence via representative models.
Recently we have  briefly  reported \cite{EPL} a novel 
approach for the  Monte Carlo computation of the critical Casimir force which 
allowed us to study the scaling behavior of $f_\Cas$ in the experimentally 
relevant cases mentioned above and to provide results for features of $f_\Cas$
which were theoretically not accessible before. 
Specifically, as follows from finite-size scaling theory 
\cite{krech:99:0,dantchev} the temperature and the geometry dependence 
of the critical Casimir  force  $f_\Cas$ per unit area $A$
 and in units of $k_BT\equiv \beta^{-1}$ can be
 expressed in terms of a universal  scaling function $\vartheta$ the form
of which  depends on the shape of the geometrical confinement, on the bulk
universality class  
of the confined medium, and on the surface universality classes of 
the confining surfaces~\cite{diehl:86:0}.
The latter are related to the boundary conditions 
(BC)~\cite{diehl:86:0,krech:99:0,dantchev} 
imposed by the surfaces on the relevant
fluctuating field, i.e.,  on the order parameter (OP) of the underlying
second-order phase transition.

Binary liquid mixtures near their demixing  points
belong to the bulk  universality class of the
three-dimensional (3D) Ising model,
whereas liquid $^4$He near the superfluid temperature of the critical end point
of the $\lambda$-line belongs to the bulk universality class of the XY model.
In the aforementioned experiments involving thin films of classical fluids,
both confining surfaces preferentially   adsorb  one or  the other 
of the two components of the
binary  mixture. This corresponds to the surface universality class
of symmetry-breaking surface fields \cite{diehl:86:0};
the sign of the surface field ($+$ or $-$) acting at the boundary 
of the  system  indicates which component
of the mixture is preferentially adsorbed.
Accordingly, $(+-)$ BC  reflect the fact
that effectively the two  surfaces  attract  different components of the liquid
mixture, whereas  $(++)$  (and, equivalently, $(--)$) 
BC correspond to the case in which
 the two  surfaces effectively attract the same component.
In the case of the colloidal suspension studied in 
Ref.~\cite{Hertlein} both surfaces
could be treated chemically 
such that  $(++)$ as well as $(+-)$ BC  have been realized.
For the wetting experiment of Ref.~\cite{pershan} the  appropriate BC are 
$(+-)$.
In the case of wetting experiments for pure 
superfluid $^4$He \cite{garcia}
the superfluid OP vanishes at both interfaces; there are no surface fields
which couple to the superfluid OP. This 
corresponds to the  symmetric Dirichlet-Dirichlet BC $(O,O)$
based on the so-called ordinary
$(O)$ surface universality class.

Due to the complexity of technical challenges as well as due to conceptual
issues like the dimensional crossover in three-dimensional films, theoretical
studies of the scaling functions of the critical Casimir forces by analytic
means have been either limited to mean-field calculations or have been
confined to the disordered phase or to BC without symmetry breaking
fields. Therefore Monte Carlo simulations offer a highly welcome tool to
overcome these shortcomings and to study, inter alia, the aforementioned
experimentally relevant universality classes within the whole temperature
range.

Our computer simulations of the critical Casimir force 
are based on
the integration scheme of free energy differences via the so-called
"coupling parameter approach"  and 
we computed the scaling
functions for the 3D Ising model with  $(++)$, $(+-)$, Dirichlet-Dirichlet $(O,O)$,  and periodic BC (PBC),
as well as for  the 3D XY model
with Dirichlet-Dirichlet $(O,O)$ and periodic 
BC. In all cases we studied the film  geometry.
The experimental data of  Refs.~\cite{pershan} and ~\cite{garcia}
turn  out to be in a good agreement with our simulation 
results which are, in addition, consistent 
with those obtained by alternative numerical approaches based on 
the computation of either the expectation value of a 
suitable 
lattice stress  tensor \cite{DK} 
for  the  3D Ising model with  
periodic BC, or of the internal energy density, followed by an
integration over the temperature~\cite{hucht}, 
for the XY model with $(O,O)$
BC. We also find good agreement with the results 
of the de Gennes-Fisher local-functional method extended to 
the Ising universality with $(++)$ BC \cite{upton,bu-08}.
%
%

The purpose of the present study  is to elucidate  the relevant details of 
the approach  used in Ref.~\cite{EPL} and to present new results
 for both the  Ising and  the XY bulk universality class in three dimensions.
In particular, we extensively discuss the important issue of 
corrections to scaling and the fitting procedure necessary to obtain
the estimates of the scaling functions $\vartheta$ from the raw MC data. 
Several functional forms of  corrections to scaling
are considered and the ensuing 
differences in the resulting scaling functions are
described. 
In particular, the estimates for the universal Casimir amplitudes at $T_c$ 
are obtained.

Our presentation is organized 
as follows:
In Sec.~\ref{sec:theory} we provide the basic theoretical background, i.e.,
the models, the critical Casimir force, and the scaling functions are defined. 
In Sec.~\ref{sec:method} we summarize our method for the computation of the 
scaling functions.
New data  for the XY model with $(O,O)$ as well as with periodic BC
are presented in Subsec.~\ref{subsec:XY}.
They have been obtained for larger lattices and with a better accuracy compared
to the results presented in Ref.~\cite{EPL}.
Discussions of the dependence of the corresponding scaling functions on the
aspect ratio of the simulation cell and of
the corrections to scaling are included. For the case of $(O,O)$ BC 
in the XY model we present the comparison
with the experimental data for wetting films of $^4$He~\cite{garcia}
 and with the MC simulation results
obtained in Refs.~\cite{DK,hucht}. Data for periodic BC in the XY model 
are compared to the available
field-theoretical  predictions  above the
bulk  critical temperature  $T_c$ ~\cite{krech:92,KD-92,DGS,GD} and to
the MC simulation data of Ref.~\cite{DK}. 
The analysis of  the 3D Ising model is 
reported in Subsec.~\ref{subsec:ising} where we present new data for the Casimir scaling function for the $(O,O)$ BC, the
aspect ratio dependence of the  Casimir scaling functions
for periodic  BC, the  determination of the universal Casimir amplitude
via the  analysis of the finite-size corrections, 
and the detailed description of the fitting procedure. 
In addition we compare our results for periodic BC in the Ising model 
with recent field-theoretical predictions  for  the behavior
of the corresponding scaling function above 
$T_{c}$~\cite{krech:92,KD-92,DGS,GD}
and with results in two dimensions (2D)
For $(++)$ and $(+-)$ BC we provide a  comparison of our data 
with the exact results in 2D \cite{ES} 
and  with mean-field predictions ~\cite{krech} as well as with 
results of the extended de Gennes-Fisher local-functional method 
applied to the case of $(++)$ BC \cite{upton,bu-08}.
The  experimental data for the scaling function obtained
from  the wetting experiments for a binary liquid mixture
 in Ref.~\cite{pershan} are compared with our MC  results for
$(+-)$ BC.
We end with a summary and conclusions in Sec.~\ref{sec:concl}.
%


\section{Theoretical background}
\label{sec:theory}
We consider the Ising and the XY model 
defined on  a three-dimensional simple cubic lattice via the Hamiltonian
\be
\label{eq:Ham}
H = - J \sum_{\la i,j \ra} {\bf s}_{i} \cdot {\bf s}_{j}, 
\ee
where $J>0$ is the spin-spin  coupling constant,
 the sum $\la i,j \ra$ runs  over all nearest neighbor pairs
of sites $i$ and $j$ on the lattice. In the Ising model,
${\bf s}_i$ has only one component $s_i \in \{ +1, -1\}$, whereas in
the XY model ${\bf s}_i$ is a two-component vector with modulus 
$|{\bf s}_i| = 1$.
Temperatures and energies are measured in units of $J$.
The inverse critical temperature is 
$\beta_c=0.2216544(3)$~\cite{RZW}
for the Ising model, whereas $\beta_c=0.45420(2)$ ~\cite{GH}  for the XY 
model.
We consider film 
geometries, i.e., lattice cells of sizes $L_{x} \times L_{y} \times L_{z}$
with $L_{x}=L_{y} \gg L_{z} \equiv L$ and $A=L_x \times L_y$,  with periodic BC
in the 
 $x$ and $y$ directions (in which the system has linear extensions $L_x$
and $L_y$).  In the $z$ direction we consider $(O,O)$ and periodic 
BC for the XY model  and fixed, $(O,O)$, and periodic BC  for the Ising model. 
The $(++)$ and $(+-)$ BC  are realized by  fixing the boundary spins 
to values  $s_i=+1$ ($+$) or $s_i=-1$ ($-$) whereas $(O,O)$ BC are realized by
free surface spins.

In a film geometry with thickness $L$ and large transverse area $A$, 
the Casimir force $f_\Cas$ per unit area $A$ 
and in units of $\kB T \equiv \beta^{-1}$ is defined as
\be
\label{eq:def}
f_\Cas(\beta,L)\equiv - \partial f^\ex/\partial L, 
\ee
where
$f^\ex(\beta,L)\equiv \beta L [f-f^\bulk(\beta)]$
is the excess free energy which depends on the type of the  BC, 
$f$  is the free energy of the film per  volume $V=LA$ %
and  $f^\bulk$ is the bulk free energy density.
From the  general theory of  finite-size scaling ~\cite{FSS} and based on
renormalization-group analyses~\cite{krech:92,KD-92}
we expect the Casimir force to  take the universal scaling form   
\be
\label{eq:scf}
f_\Cas(\beta,L)=L^{-d}\vartheta\left(\tau(L/\xi_0^+)^{1/\nu}\right)
\ee
where the scaling function $\vartheta(x)$ depends on the 
spatial dimension $d$
and on  the BC. Here 
$\tau=
(\beta_{c}-\beta)/\beta = (T-T_{c})/T_{c}$ is the reduced 
temperature and $\xi=\xi_{0}^{\pm} |\tau|^{-\nu}$ is the {\it bulk}
correlation length which controls the spatial 
exponential decay of the two-point correlation function. 
The critical exponent $\nu$ equals
$0.6301(4)$ and $0.662(7)$ for the  Ising and the XY
bulk universality class in three dimensions, respectively~\cite{PV};
$\xi_0^\pm$ are nonuniversal amplitudes above $(+)$ and below $(-)$ $T_c$ with
$\xi^+_0=0.501(2)$~\cite{RZW}
for the Ising model on the simple cubic lattice, whereas 
$\xi_0^+=0.498(2)$~\cite{GH}  for the XY model.
The values of $\xi_0^+$ quoted here
refer to the amplitude of the 
{\it second moment} correlation length $\xi_{\mathrm{2^{nd}}}$;
however,
$\xi/\xi_{\mathrm{2^{nd}}} \simeq 1$ for $\beta < \beta_c$ for both the
Ising and the XY model~\cite{PV,GH}.

At $T=T_c$ the scaling 
function reduces to the {\it  universal Casimir amplitude}
$\vartheta\left(0\right)\equiv  (d-1)\Delta$, 
which has been extensively  studied
in the literature (see, e.g., 
Refs.~\cite{krech:92,KD-92,krech:99:0,dantchev,upton,DGS}). 
Determining the whole temperature
dependence of the scaling function and its dependence on the spatial dimension
$d$ is a much more challenging task.

For the Ising  universality class with $(O,O)$, $(++)$, and  $(+-)$ BC in the
film geometry theoretical results are available 
in  $d=2$ from the exact diagonalization
of the transfer matrix~\cite{ES} and 
in $d> 4$ from mean-field theory~\cite{krech}.
In $d=3$ theoretical results are   available for $T\ge T_c$ 
and periodic BC investigated both  by MC
simulations (at $T_c$)~\cite{DK} and by field-theoretical
methods~\cite{krech:92,KD-92,DGS,GD} 
as well as  for Dirichlet~\cite{krech:92,KD-92}, 
von Neumann BC~\cite{krech:92,KD-92}, and Robin BC \cite{SD:08}
investigated by field-theoretical methods.
Recently, the extended de Gennes-Fisher local-functional method
has been applied in order to study  the case of $(++)$ BC within the full
temperature range~\cite{upton,bu-08}. 

For the bulk universality class of the  XY model in   film geometry 
with $(O,O)$ BC theoretical results for  the Casimir force scaling function
are available in $d=3$. They include  
field-theoretical calculations 
for temperatures $T\ge T_c$~\cite{krech:92,KD-92}
and numerical results from  MC simulations~\cite{EPL,hucht}.
In the low temperature limit the specific features of the 
superfluid $^4$He were taken into account in Ref.~\cite{kardar:04} and
the contribution to the Casimir force resulting from the capillary-wave like
fluctuations on the surface of $^4$He wetting films 
was determined. For $T < T_c$, which corresponds to temperatures
below the superfluid-normal fluid transition temperature $T_{\lambda}$
of the $\lambda$ transition, certain qualitative features 
of the Casimir scaling function have been recently understood within
the framework of  the Landau-Ginzburg mean-field theory~\cite{LGW-MF,LGW-MF1}.

For large areas  $A$, the total free energy
$F(\beta,L,A)$ of the confined system can be written as 
\be
\label{free_en1}
F(\beta,L,A) \equiv A L f = A [L f^\bulk(\beta) + \beta^{-1} f^\ex(\beta,L)].
\ee
The quantity $f^\ex$ contains two $L$-independent surface  contributions
in addition to the finite-size contribution
$f^\ex(\beta,L)-f^\ex(\beta,\infty)$ the $L$-dependence of which 
gives rise to the effective Casimir force.
On a lattice ($\,\hat{}\,$), the derivative in Eq.~(\ref{eq:def})
is replaced by a finite difference and $f_\Cas(\beta,L)$ is given by 
\be 
\label{eq:force}
\hat f_\Cas(\beta,L-\frac{1}{2},A)\equiv - \frac{\beta \Delta F(\beta,L,A)}{A}
+ \beta f^\bulk(\beta)\,,
\ee
where $\Delta F(\beta,L,A)= F(\beta,L,A)-
F(\beta,L-1,A)$.
One can consider different definitions of the lattice derivative than the one we have implemented in Eq.~\reff{eq:force}. Different
choices give rise to different corrections to the leading behavior
of the Casimir force scaling function. 
%
%

\section{Method}
\label{sec:method}

\subsection{Computation of free energy differences}
\label{subsec:fe}
Monte Carlo methods are generally not efficient for the computation of
quantities, such as the free energy $F$, which cannot be expressed as ensemble
averages. Nevertheless, free energy differences, such as $\Delta F(\beta,L,A)$
we are interested in, can be cast in such a form via the
so-called ``coupling parameter approach'' (see, e.g., Ref.~\cite{Mon}). 
This is a viable alternative to the method
used in Ref.~\cite{DK} 
in which a suitable lattice stress tensor has been defined in such a way 
that its ensemble average renders $\Delta F$. So far, however, this 
latter method is
only applicable for periodic BC.
%
%
\begin{figure}
\includegraphics[height=5.5cm,width=0.45\textwidth]{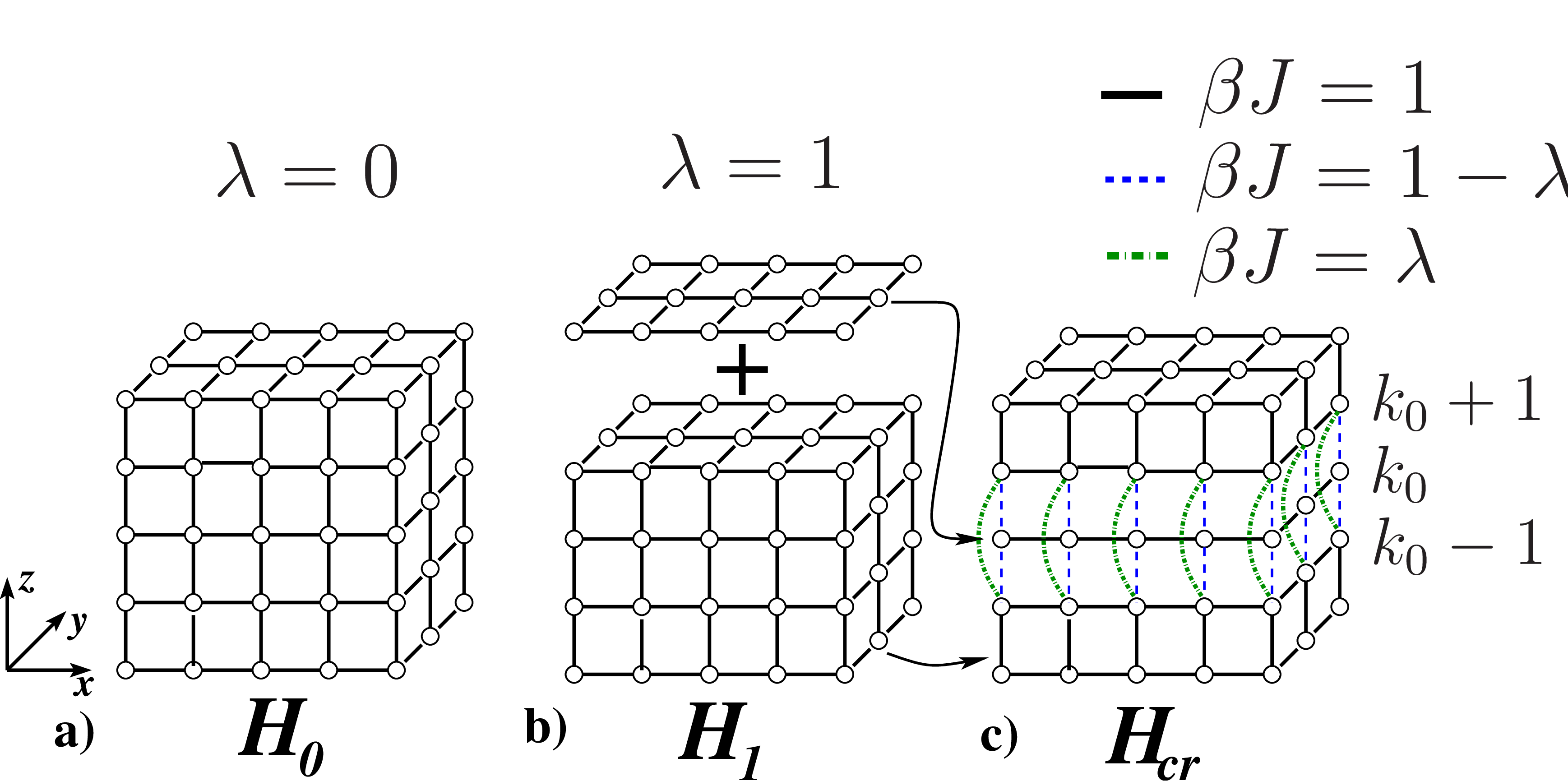}
\caption{%
Bond arrangement for the computation of the free energy 
difference in Eq.~\protect{\reff{DF}} (see main text). The crossover
Hamiltonian $H_{cr}$ (c) belongs to a system which interpolates between those
described by the Hamiltonians $H_0$ (a) and $H_1$ (b). 
}
\label{fig:fig1}
\end{figure}
%

%
If one is interested in the Monte Carlo computation of
the difference $F_1-F_0$ between the free
energies  $F_i = -\frac{1}{\beta}\ln \sum_{\C} \exp(-\beta H_i)$ ($i\in\{0,1\}$) of two 
lattice models ${\mathcal M}_0$ and ${\mathcal M}_1$ 
with the same configuration space
$\C$ but different Hamiltonians $H_0$ and $H_1$, respectively, 
it is convenient to introduce an ``interpolating'' 
system ${\mathcal M}_{\rm cr}(\lambda)$  with the {\it cr}ossover 
Hamiltonian
\be
H_{\rm cr}(\lambda)=(1-\lambda)H_0 +\lambda H_1,
\label{Hcr}
\ee
where $\lambda\in[0,1]$, and again the same configuration space $\C$.
As a function of the coupling
parameter $\lambda$, $H_{\rm cr}(\lambda)$ 
interpolates between $H_0$ and $H_1$ as $\lambda$
increases from 0 to 1 and accordingly the free energy 
$F_{\rm cr}(\lambda) =  -\frac{1}{\beta}\ln \sum_{\C} \exp(-\beta
H_{\rm cr}(\lambda))$ of ${\mathcal M}_{\rm cr}(\lambda)$ 
interpolates between $F_0$ and $F_1$. 
The difference $F_1-F_0$ can be trivially expressed as
$F_1-F_0 = \int_0^1\rmd \lambda \, F'_{\rm cr}(\lambda)$ where
$F'_{\rm cr}$ is the derivative of $F_{\rm cr}(\lambda)$ with respect to the
coupling parameter:
\be
\frac{\rmd F_{\rm cr}(\lambda)}{\rmd \lambda} =
\frac{\sum_\C (H_1-H_0) \rme^{-\beta H_{\rm cr}(\lambda)}}{\sum_\C \rme^{-\beta H_{\rm cr}(\lambda)} }
=\la\Delta H\ra_{{\mathcal M}_{\rm cr}(\lambda)}\,,
\ee
which takes the form of the canonical ensemble average $\la \ldots
\ra_{{\mathcal M}_{\rm cr}(\lambda)}$ 
of $\Delta H \equiv H_1 - H_0$ and therefore it can be
efficiently computed via MC simulations of the lattice model ${\mathcal
  M}_{\rm cr}(\lambda)$.
As a result one can conveniently 
express the difference in free energies as an integral
over canonical averages (see, e.g., Ref.~\cite{Mon}):
\be
F_1 - F_0 = \int_0^1\!\!\rmd\lambda \, \la\Delta
H\ra_{{\mathcal M}_{\rm cr}(\lambda)} \, .
\label{DF}
\ee

According to Eq.~\reff{eq:force}, 
the Casimir force we are interested in is related
to the difference $\Delta F(\beta,L,A)$ 
between the free energies $F(\beta,L,A)$ and $F(\beta,L-1,A)$ of the
same model on two lattices with different numbers of sites and
therefore different configuration spaces. 
In order to apply the method described above for the computation of
$\Delta F(\beta,L,A)$ one identifies the model ${\mathcal M}_0$, its
Hamiltonian, and the associated 
configuration space $\C$ with the corresponding ones of
the model we are interested in on the lattice $A\times L$, 
as depicted in Fig.~\ref{fig:fig1}(a), so that 
$F_0(\beta,L,A) = F(\beta,L,A)$.
The final system ${\mathcal M}_1$ has to be chosen such that it has the same
configuration space $\C$ as ${\mathcal M}_0$. This is achieved 
by adding 
to the model on the lattice $A \times (L-1)$ -- for which we want to compute
the free energy  $F(\beta,L-1,A)$ -- a two-dimensional lattice of size
$A$ with suitable degrees of freedom and lateral periodic BC
(see Fig.~\ref{fig:fig1}(b)).
The Hamiltonian $H_1$ of ${\mathcal M}_1$ is defined such that the
added layer does not interact with the remaining part of the system and
therefore $F_1(\beta,L,A) = F(\beta,L-1,A) + F_\mathrm{2D}(\beta,A)$, where 
$F_\mathrm{2D}(\beta,A)$ 
is the free energy of the isolated two-dimensional layer. 
This layer can be thought of as the one at position
$k_0\in\{1,2,\ldots,L\}$  (along the $z$-direction) in the model 
${\mathcal  M}_0$ which then decouples from the rest of the lattice 
upon passing from $\lambda=0$ to
$\lambda=1$, i.e., from Fig.~\ref{fig:fig1} (a) to (b). 
The resulting crossover Hamiltonian 
$H_{\rm cr}(\lambda)$ (see Eq.~\reff{Hcr})
 additionally depends on the original 
position $k_0$ of the extracted layer.
In particular, in the three-dimensional 
models we are mainly interested in, the fluctuating degrees of
freedom are one- (Ising) or two-component (XY) vectors ${\bf 
s}_{x,y,z}$ --- where $i = (x,y,z)$ specifies the lattice site --- which
interact only with their nearest neighbors on the same
lattice, with a coupling strength $J=1$ (indicated by solid bonds in
Figs.~\ref{fig:fig1} (a) and (b); $J$ is absorbed into $\beta$). 
For them one explicitly finds
\be
\begin{split}
\Delta H \equiv H_1 - H_0 =& -\sum_{x,y}
({\bf s}_{x,y,k_0-1}\cdot {\bf s}_{x,y,k_0+1} \\
-& 
{\bf s}_{x,y,k_0-1}\cdot {\bf s}_{x,y,k_0}- {\bf s}_{x,y,k_0}\cdot 
{\bf s}_{x,y,k_0+1}) \;.
\end{split}
\label{eq:deltaH}
\ee
The resulting $H_{\rm cr}(\lambda) = H_0 + \lambda \Delta H$ 
is characterized by the coupling
constants depicted in Fig.~\ref{fig:fig1}(c).
The free energy difference 
$\Delta F$ (see
Eqs.~\reff{eq:force} and~\reff{DF}) can be finally expressed as
\be
\Delta F(\beta,L,A) = - I(\beta,L,A) + F_\mathrm{2D}(\beta,A)
\ee
where $I(\beta,L,A) = \int_0^1\rmd\lambda \langle\Delta H\rangle_{{\mathcal
    M}_{\rm cr}(\lambda)}$.
Note that $H_{\rm cr}(\lambda)$
(see Fig.~\ref{fig:fig1}(c)), $\Delta H$ (see Eq.~\reff{eq:deltaH}), %
and therefore $\la \Delta H\ra_{H_{\rm cr}(\lambda)}$ depend on the
value of $k_0$ whereas $I(\beta,L,A)$ is actually independent of it, 
as long as the boundary conditions are not affected by the extraction of the
$k_0$-th layer as $\lambda$ varies between 0 and 1. For fixed and open BC in
the $z$-direction this requires  $k_0\neq 1$, whereas for PBC there is no such
a restriction on $k_0$ and 
$\la \Delta H\ra_{H_{\rm cr}(\lambda)}$ is actually  independent of it.
In our simulations we have chosen $k_0 = L/2$. 

Once $\Delta F(\beta,L,A)$ has been computed, one has still to subtract 
$f^\bulk(\beta)$ from it (see Eq.~\reff{eq:force}), in order to determine the
Casimir force in a film of assigned thickness $L-1/2$.
However,
the accurate computation of the bulk free energy density $f^\bulk(\beta)$ 
is a numerical problem by itself and extracting it from finite-size data
requires a very accurate analysis. In order to avoid this complication
 in the computation of
the Casimir force, it is convenient to consider the {\it difference} between
the forces acting in slabs of thicknesses $L_1$ and $L_2>L_1$:
\be
\begin{split}
\Delta  &\hat f_\Cas(\beta,L_1,L_2,A)\\
 &\equiv 
\hat f_\Cas(\beta,L_{1}-\frac{1}{2},A)-\hat
f_\Cas(\beta,L_{2}-\frac{1}{2},A)\\
&= \frac{\beta}{A}[I(\beta,L_1,A)-I(\beta,L_2,A)]
\end{split}
\label{eq:deltaC}
\ee 
in which the contributions of both 
$f^\bulk(\beta)$ and $F_\mathrm{2D}(\beta,A)$ actually 
cancel. 
Accordingly, the procedure to calculate the scaling function of the Casimir
force consists of the following steps:
(1) For a given geometry $L\times A$ and temperature $\beta^{-1}$,
via MC simulations  we compute 
the ensemble averages $\la \Delta H\ra_{H_{\rm cr}(\lambda)}$ 
for different values of $\lambda\in\{\lambda_1, \ldots \lambda_N\}$. 
(2) On the basis of these $N$ values we 
calculate the integral $I(\beta,L,A)$ in Eq.~\reff{DF} via
numerical integration.  
(3) These computations are repeated for different
sizes $L$, $A$, and temperatures $\beta^{-1}$, yielding numerical estimates
for $\Delta \hat f_\Cas(\beta,L_1,L_2,A)$ (see Eq.~\reff{eq:deltaC}). 
(4) The scaling function $\vartheta$ in
Eq.~\reff{eq:scf} is retrieved from the numerical data for $\Delta \hat
f_\Cas$ as described below.
The results presented in Sec.~\ref{sec:res} have been obtained by using the
Simpson integration method with $N=20$ mentioned above in step (2) and by
using pairs of geometries
$(L_1,L_2) = (L,2 L)$ with $L=13, 16, 20$ for the Ising model and $L=10,15,20$
for the XY model,  as introduced in step (4) and for fixed 
aspect ratios  $\rho \equiv L/\sqrt{A}$.
(The motivation for our choice $L_2 = 2 L$ will be provided in the following
subsection.)
The  method of Ref.~\cite{hucht}  takes advantage of the possibly
available numerical knowledge of the bulk energy density $u^\bulk$
of the model of interest whereas here the analogous information
on the bulk free energy $f^\bulk$ is not required for the determination of
the Casimir scaling function,
 making our approach applicable also to cases in which there
is no detailed knowledge of $u^\bulk$ and $f^\bulk$.

\subsection{Determination of the scaling function}
\label{subsec:determ}

The scaling function $\sf$ of the Casimir force can be extracted
from the temperature dependence of $\Delta \hat f_\Cas(\beta,L_1,L_2,A)$, for
fixed $L_{1,2}$ and $A$, by using the fact that $\hat f_\Cas$ in
Eq.~\reff{eq:deltaC} scales according to Eq.~\reff{eq:scf} for large
$L_{1,2}$ and $A$.
In order to highlight these 
scaling properties it is convenient to introduce the 
quantity
\be
\begin{split}
&g(y;L_1,L_2,A)\equiv \left(L_{1}-1/2 \right)^{d} \times \\
&\quad\quad\quad\quad \Delta \hat f_\Cas(\beta = \beta(y;L_1),L_1,L_2,A)\,,
\end{split}
\label{eq:defg}
\ee
as a function of $y$, where $\beta(y;L_1) \equiv  \beta_c/[1+ y
(L_1-1/2)^{-1/\nu}]$. According to Eq.~\reff{eq:scf} and with
$\tau=(\beta_c-\beta)/\beta$, $g$ is expected to scale as
\be
g(y;L_1,L_2,A) = \sfhb(y) - \alpha^{-d} \sfhb(\alpha^{1/\nu} y)\,,
\label{eq:implicit}
\ee
where $\alpha = (L_2-1/2)/(L_1-1/2)$ is the width ratio, and $\sfhb(y)$
is the Monte Carlo estimate of $\sfb(y) \equiv \sf(y/(\xi_0^+)^{1/\nu})$; here
$d=3$. 
Note that, even though $\sfb$ is independent of this geometrical realization
of the simulation cell, $\sfhb$ might
depend on it via $A$ and $L_{1,2}$ due to corrections to scaling. 
For a given pair of geometries $L_1\times A$ and $L_2\times A$, 
the available Monte Carlo data for  $\Delta \hat
f_\Cas(\beta,L_1,L_2,A)$ at different temperatures allow one to determine
$g(y;L_1,L_2,A)$ for a discrete set of values of $y$. 
In order to determine $\sfhb(y)$ from the numerical data for  $g(y;L_1,L_2,A)$
with fixed $L_{1,2}$ and $A$, one can solve 
Eq.~\reff{eq:implicit} iteratively. 
One can expect (see below) that this
yields a solution for $L_2>L_1$, i.e., $\alpha > 1$ together with the property 
$\sfhb(|y|\rightarrow\infty) \rightarrow 0$ (which holds 
apart from $T<T_c$ in the XY
model).
As a first approximation of  the actual $\sfhb(y)$ one takes 
$\sfhb_{0}(y) \equiv g(y;L_{1},L_{2},A)$, which can be improved by taking into
account that Eq.~\reff{eq:implicit} yields
$\sfhb(y) = \sfhb_0(y) + \alpha^{-d}\sfhb(\alpha^{1/\nu}y) \simeq
\sfhb_0(y) + \alpha^{-d}\sfhb_0(\alpha^{1/\nu}y)$. Accordingly, a better 
approximant $\sfhb_1(y)$ is provided by
\be
\sfhb_1(y) = \sfhb_0(y) + \alpha^{-d}\sfhb_0(\alpha^{1/\nu}y). 
\label{eq:implicit1}
\ee
The values of $\sfhb_0$ at the point $\alpha^{1/\nu} y$, for which no MC data
might be available, are obtained by cubic spline interpolation of
the available ones. 
In Eq.~\reff{eq:implicit1} one can replace $\sfhb_0$ by using
Eq.~\reff{eq:implicit}, yielding 
$\sfhb_{1}(y)=\sfhb(y)-\alpha^{- 2d}\sfhb(\alpha^{2/\nu} y) \simeq \sfhb(y)-\alpha^{- 2d}\sfhb_{1}(\alpha^{2/\nu} y)$, which indicates
how the approximant $\sfhb_{1}(y)$ can be improved by 
introducing $\sfhb_{2}(y)=\sfhb_{1}(y)+\alpha^{-2 d}\sfhb_{1}( \alpha^{2/
\nu} y) = \sfhb(y)-\alpha^{- 4d}\sfhb(\alpha^{4/ \nu} y)$. This expression can
in turn be used to further improve the approximant along the same lines. 
The resulting iterative procedure yields a sequence of approximants
\be
\sfhb_{k\ge1}(y)=\sfhb_{k-1}(y)+\alpha^{-2^{k-1} d}\sfhb_{k-1}( 
\alpha^{2^{k-1}/\nu} y),
\label{eq:approximants}
\ee
which converges very rapidly because the correction
to the $k$-th approximant is 
of the order of $\alpha^{-2^{k-1}d}$, i.e., exponentially small 
in $2^k$ and, in addition, $\sfhb(y)$ is generally 
expected to decay exponentially for large $|y|$.
With $\alpha \simeq 2$, already for $k=5$ one has $\alpha^{-2^{k-1}d}\simeq
3.5 \times 10^{-15}$ in three dimensions ($d=3$).
The choice of $\alpha\simeq L_2/L_1$ is a compromise between two competing 
aims: a small value reduces the sizes of the
lattices required for the computation of $\hat f_\Cas$ but on the other hand
it decreases  the accuracy of a given approximant in determining $\hat \sfb$.
With our choice of geometries $(L_1,L_2)=(L, 2L)$, one has 
$\alpha \simeq 2$ and a very good approximation
of $\sfhb(y) \equiv \sfhb_{k\to\infty}(y)$ is already provided by
$\sfhb_5(y)$. 
%
%

\subsection{Details of the MC simulations and test of the method}
\label{subsec:details}
In order to compute  the canonical average 
$\la \Delta H\ra_{{\mathcal M}_{\rm cr}(\lambda)}$ we use a
hybrid MC method  which is a suitable mixture of Wolff 
and  Metropolis algorithms~\cite{LB}.
Specifically, for the Ising model each hybrid MC step consists of 
four flips of a Wolff cluster according to the Wolff algorithm, 
typically followed by $3 A$  
attempts to flip a spin $s_{x,y,z}$ with $z\in \{k_0-1,k_0,k_0+1\}$,
which are accepted according to the Metropolis rate~\cite{LB}. 
An analogous
method, with a suitable implementation of Metropolis and Wolff algorithms, 
has been used for the XY model~\cite{GH}, i.e.,
a flip of a Wolff cluster according to the Wolff algorithm is 
typically followed by the implementation of moves according to the
 Metropolis algorithm  ~\cite{LB}.

In order to test the program we have computed numerically $g(y,5,10,9)$  
as a function of $y$ for the Ising model on a lattice   
$3 \times 3 \times L$ with periodic, $(++)$, and $(+-)$ BC, 
finding perfect agreement with the result of the analytic
calculation based on the transfer-matrix method.
%
%
%

\subsection{Corrections to scaling}
\label{subsec:corr}

Finite-size scaling  is known to be  valid asymptotically for 
large lattices  and small values of $\tau$, i.e., a large correlation length
$\xi$~\cite{FSS}. Away from
the asymptotic regime corrections to the leading (universal) scaling behavior
become relevant. These non-universal 
corrections affect both the scaling variables
and the scaling functions and depend on the details of the model
as well as on the geometry and the boundary conditions~\cite{privman,luck}.
Renormalization-group analyses reveal that 
there is a whole variety of sources of corrections which arise
from bulk, surface, and finite-size effects~\cite{FSS}. 
For the limited thicknesses $L$ of the lattices we 
investigated with our MC simulations 
it is necessary to take  corrections to scaling  into account in order to
obtain data collapse~\cite{hucht,EPL}.
In the present case,
the finite-size scaling variable $\tau (L/\xi_0^+)^{1/\nu}$ (in the
following associated with
the reduced temperature $\tau$) is expected to
acquire a leading correction of the form 
\be
\label{eq:xcorr}
x \equiv \tau (L/\xi_0^+)^{1/\nu} (1 + g_\omega L^{-\omega}),
\ee
where $\omega$ is the  leading correction-to-scaling
exponent in the bulk which takes the values  $0.84(4)$ and 
$0.79(2)$~\cite{PV} for the three-dimensional Ising and XY universality
class, respectively.
Corrections to the scaling behavior of the critical
Casimir   force $\hat f_\Cas$ are expected to be of the form 
\be
\begin{split}
\label{eq:sfhcorr}
\hat f_\Cas(\beta,L,A)
& = L^{-d} \sfh(x,L^{-\omega'}) \\
&\simeq L^{-d} \sf(x) [1 +
L^{-\omega'} \phi(x)  + \ldots],
\end{split}
\ee
for $ L\gg 1$, where the exponent $\omega'$  controls the leading
corrections to the scaling behavior of the lattice estimate $\hat f_\Cas$. 
Its value is determined by that irrelevant surface or bulk perturbation of the
Hamiltonian $H$ which has the smallest scaling dimension and which also
affects $\hat f_\Cas$. In the generic bulk case one has 
$\omega' = \omega$. But its
value can be suitably increased (so that the influence of the corrections is
reduced) by using improved Hamiltonians and
observables, which can also serve as representatives of the same universality
class. This is described in detail in Ref.~\cite{PV}. 
In the presence of surfaces, irrelevant surface perturbations might yield 
$\omega'<\omega$, but we are not aware of either 
theoretical or numerical studies of this issue. 
In addition, for small lattice sizes,
next-to-leading corrections to scaling
might also be of relevance. 
If $\omega'>1/2$,
these corrections are generically provided by analytic terms $\sim L^{-1}$
(even though they might be absent in some quantities). 
The interplay between
the leading and next-to-leading corrections (especially if they are sizable)
might result in an effective exponent $\omegaeff$.
The current accuracy of our Monte Carlo data and the relatively small range of
sizes $L$ investigated here do not allow a reliable
determination of $\omega'$ and $\phi(x)$. In particular it will turn out that
the corrections to scaling are quite well captured by assuming 
$\phi(x)\simeq g_2$, i.e., a constant within the range of the scaling variable
we have investigated, and an effective exponent $\omegaeff$ for the size
dependence.

In the discussion of the expected scaling behavior of $\hat f_\Cas$ we have
assumed that the aspect ratio  $\rho \equiv L/\sqrt{A}$  is small enough
(i.e., $\rho \ll 1$) so that the scaling behavior in Eq.~\reff{eq:scf} 
holds, which
formally corresponds to the limit $A\rightarrow\infty$. 
On the other hand, the actual Monte Carlo simulations have been 
performed on
lattices with small but non-zero $\rho$ and therefore possible additional, 
$\rho$-dependent corrections have to be taken into account in order to be 
able to extrapolate our results to the limit $\rho\to 0$. 
The numerical results in
Ref.~\cite{MN} on the (universal) $\rho$-dependence of the Casimir amplitude
$\sf(0,\rho)$  (see Eq.~\reff{eq:scf}) of the three-dimensional XY
model with periodic and free
boundary conditions suggest $\sf(0,0)\simeq \sf(0,\rho) (1 + r
\rho^2)$ for
$\rho \lesssim 0.5$, where $r$ is a constant. 
(This is confirmed also by the analysis in Ref.~\cite{hucht}.)
In what follows we assume that this dependence on $\rho$ carries over to the 
whole scaling function so that 
$\sf(x,0)\simeq \sf(x,\rho)(1 + \phi_2(x) \rho^2)$.
Although the amplitude $\phi_2(x)$ of the correction might depend on the
scaling variable $x$ (and possibly on $L^{-\omega'}$), 
we shall assume that $\phi_2(x) \simeq r_2$, i.e.,
a constant at least within 
the range of values of the scaling variable $x$ which 
is studied in the present analysis. 
On the same footing, we expect a quadratic 
$\rho$-dependence of the finite-size scaling variable
 $x(\rho) \simeq x(0)(1+ r_1 \rho^2)$  associated with the reduced 
temperature
$\tau$, where $r_1$ is a constant and $x(0)$ is given by Eq.~\reff{eq:xcorr}.
Taking into account all these corrections, we identify
\be 
\label{eq:xy_x_fit}
x=\tau \left( \frac{L}{\xi_0^{+}} \right)^{\frac{1}{\nu}} 
(1+g_{\omega}L^{-\omega})(1+r_{1}\rho^{2}),
\ee
as the finite-size scaling variable,
in terms of which the expected scaling behavior of $\hat f_\Cas$
is given by
\be
\label{eq:xy_f_fit}
\hat f_{\Cas}(\beta
,L,A)=L^{-d}(1+ g_2 L^{-\omegaeff}) (1+r_{2}\rho^{2})^{-1}\sf(x) .
\ee
We shall aim at fixing the
non-universal constants $r_{1,2}$, $g_\omega$,  and $g_2$, 
which generally 
depend on the boundary conditions,
in such a way that the data collapse of the available Monte Carlo 
data is optimal.

In most of the cases considered below, the
accuracy of the data and the range of sizes $L$ investigated 
do not allow for the
reliable determination of both the amplitude $g_2$ and the exponent
$\omegaeff$ of the correction. Therefore we fix $\omegaeff = 1 \simeq
\omega'$, which actually leads to a reasonably good data collapse 
within the considered range
of the scaling variable.

In the absence of corresponding
  dedicated theoretical and numerical analyses, there is no
a priori reason why one should prefer the use of a specific form of 
corrections to scaling, because all of them amount to an effective way of
accounting for
these corrections. Accordingly, adopting a pragmatic approach, 
we shall choose that form which leads to the best data
collapse or to the best fit. 
Specifically, we use the following 
functional forms of corrections to  scaling: 
\be
\begin{split}
\label{eq:xy_f_fit1}
& \mbox{case (i):} \\
& \hat f_{\Cas}(\beta
,L,A)=L^{-d}(1+g_{1}L^{-1})^{-1} (1+r_{2}\rho^{2})^{-1}\sfh
\left( x\right),
\end{split}
\ee
\be
\begin{split}
\label{eq:xy_f_fit2}
& \mbox{case (ii):} \\
&\hat f_{\Cas}(\beta
,L,A)=L^{-d}(1+g_{2}L^{-1}) (1+r_{2}\rho^{2})^{-1}\sfh(x),
\end{split}
\ee
\be
\begin{split}
\label{eq:xy_f_fit3}
& \mbox{case (iii):} \qquad \qquad \\
& \hat f_{\Cas}(\beta
,L,A)=L^{-d}(1+g_{3}L^{-\omegaeff})\sfh
\left( x\right), \qquad \qquad
\end{split}
\ee
and 
\be
\begin{split}
\label{eq:fit_new}
& \mbox{case (iv):} \\
& \hat f_{\Cas}(\beta
,L,A)=L^{-d}\frac{(1+{\tilde g}_{1}L^{-1})}{(1+{\tilde g}_{2}L^{-1})}(1+r_{2}\rho^{2})^{-1}\sfh(x).
\end{split}
\ee
Case (i), (ii),  and (iv) become all equivalent for large lattice sizes $L$.
On the other
hand, for smaller lattice sizes, they lead do different estimates. The
coefficients $g_1, g_2, g_3$ and $\tilde g_1, \tilde g_2$ 
are determined in such a way as to
optimize the data collapse in the resulting estimate for $\sf(x)$ (see below). 
The factor of the form (iv), with two
fitting parameters, will be considered 
only if data corresponding to several different values of $L$
are available, so that the resulting  estimates for $\tilde g_1$ and 
$\tilde g_2$ are reliable.
Case (iii) of corrections to scaling works well for the XY and the Ising
model with periodic BC.
In cases in which corrections to scaling are not small, 
the ansatz used
for their dependence on $L$ might lead to a biased estimate of the scaling
function $\sf(x)$. 

In order to highlight and assess 
the relevance of the different kinds of corrections, we
present in the following sections also the MC data for the function
$g(y;L,2L,A)$ which is the primary quantity determined by our MC simulation and
from which the scaling function $\sfh(x)$ is eventually obtained according to
the procedure described in Subsec.~\ref{subsec:determ}. In the absence of
corrections to scaling, data for $g$ (see Eq.~\reff{eq:defg}) with
$L_2=2L_1$, $L_1=L$,
as a function of $y =
(\beta_c/\beta-1)(L-1/2)^{1/\nu} = \tau (L-1/2)^{1/\nu}$ 
with fixed $\rho = L/\sqrt{A}$ but different
sizes $L$  should collapse on a single master curve, which, however, it is not
always the case (see, c.f., Fig.~\ref{fig:xy_gy}(a)).
In order to account for the corrections to scaling we proceed as follows:
First, for fixed values of $L$ and $A= (L/\rho)^2$ we determine the Monte Carlo
data for $g$ (see Eq.~\reff{eq:defg}) 
for different values of the inverse temperature $\beta$. Second, from the
plot of $g$ as a function of the rescaled reduced temperature  $y =
\tau (L-1/2)^{1/\nu}$, i.e., from $g(y;L,2L,A)$, we determine the
estimate of the scaling function $\sfhb(y)$, according to the procedure 
described in Subsec.~\ref{subsec:determ}. 
This procedure is repeated for the
different geometries considered in each case. 
Because of corrections to
scaling and corrections due to $\rho\neq 0$, the resulting estimates
$\sfhb(y)$ actually depend on the specific values of $L$ and $A=(L/\rho)^2$, 
i.e., $\sfhb = \sfhb(y;L,\rho)$. 
In order to extract the asymptotic limit $\sfb$ of the scaling function of the
Casimir force from the lattice estimate $\sfhb$, we account for corrections
in accordance with Eqs.~\reff{eq:xy_x_fit} and~\reff{eq:xy_f_fit} 
(with possibly different forms for the
$L$-dependent corrections, see Eqs.~\reff{eq:xy_f_fit1}--\reff{eq:fit_new}), 
which involve several fitting parameters.
In those cases in which  we apply corrections to scaling 
due to the aspect ratio dependence
of  the function $g(y;L,2L,A)$ (which turns out to be the case only
for the XY model),
the actual fitting procedure we shall use is divided into two steps.

In the first step we fix the value of $L$ ($L=10$ for the XY model) 
and consider data corresponding to different aspect
ratios  $\rho$ ($\rho^{-1}=4,5,6,8,10$ for XY). 
The parameters 
$r_1$ and $r_2$ are therefore determined such that the 
data for $(1 + r_2 \rho^2)^{-1} \sfhb(y;L,\rho)$ ($\propto
\sfb(x)$ for fixed $L$) as a
function of $y (1 + r_1 \rho^2)$ ($\propto x$, see Eq.~\reff{eq:xcorr}, 
for fixed $L$) yield the best data collapse onto a single curve which ideally
corresponds to the scaling function in the limit $A\rightarrow \infty$, but
which is still affected by $L$-dependent corrections to scaling.
(This procedure actually assumes that, according to Eqs.~\reff{eq:xy_x_fit}
and~\reff{eq:xy_f_fit}, $r_1$ and $r_2$ do not depend on $L$.) 

In the second  step we fix the value of $\rho$ ($\rho = 1/6$  for both XY
and  Ising) and we determine $g_\omega$ and $g_{2}$ (or $g_1$ 
or both ${\tilde g}_{1}$  and ${\tilde g}_{2}$, 
depending on the specific form assumed for the
corrections) in such a way that the data for
$(1+ g_2 L^{-\omegaeff})\sfhb(y;L,\rho)$ ($\propto
\sfb(x)$ for fixed $\rho$) as a function of $y (1 + g_\omega L^{-\omega})$
($\propto x$, see Eq.~\reff{eq:xcorr},  for fixed $\rho$) 
yield the best data collapse onto a single curve.
(This procedure actually assumes that, according to Eqs.~\reff{eq:xy_x_fit}
and~\reff{eq:xy_f_fit}, in a first approximation corrections to scaling 
do not depend  on
$\rho$.) 
The details of the fitting  procedure are described  in 
Appendix~\ref{sec:size_corr}.
Our final numerical estimate of the scaling function $\sf(x)$ of the Casimir
force is then provided by the curve which result from plotting
$(1+g_2 L^{-\omegaeff})(1 + r_2 \rho^2)^{-1}\sfhb(y;L,\rho)$ 
(or equivalent forms as given by the cases (i)-(iv)) as a
function of $y (1 + g_\omega L^{-\omega}) (1 + r_1 \rho^2)/(\xi_0^+)^{1/\nu}
\equiv x$, where the fitting parameters have been fixed according to the
procedure described above.

Finally it is worthwhile to keep in mind that besides the common corrections
to the leading critical behavior in experimental data, the available
experimental results for critical Casimir forces contain an additional source
of corrections in that the thickness $L$ of the (wetting) films is
definition-dependent up to a microscopic length
$\ell_0$~\cite{DSD:07}. Accordingly, only the leading term is universal whereas
the correction $\sim \ell_0/L$ is even definition-dependent. Moreover, also
the relation between the experimental values $L_{\rm exp}$ and the theoretical
values $L_{\rm theo}$ suffers from the same kind of uncertainty.   

\section{Results}
\label{sec:res}

In this section we summarize the numerical results for the scaling function of
the critical Casimir force within the three-dimensional XY 
(Subsec.~\ref{subsec:XY})
and Ising (Subsec.~\ref{subsec:ising}) universality classes with different
boundary conditions. As mentioned in
the Introduction, the former are relevant for the interpretation of the
experiments with wetting films of $^4$He~\cite{garcia}, whereas the latter
apply to 
the case of classical binary mixtures~\cite{pershan,Hertlein}.
In  most of the  presented plots the size of the symbols are of the order
of the statistical error. In these cases  
 the corresponding error bars are not shown in the figures.

\subsection{XY model} 
\label{subsec:XY}
%
%

For the simulations of the XY model we have considered films of thicknesses
$L=10, 15$, and $20$, and transverse areas $A = L_x \times L_y= 6L \times 6L$
corresponding to an 
aspect ratio $\rho = L/\sqrt A=1/6$.
 At the boundaries in the $x$- and $y$-directions we impose
periodic BC, whereas in the $z$-direction we consider either free surface
spins, corresponding to the $(O,O)$ universality class, or periodic BC. 
%
\begin{figure}
\includegraphics[height=8cm,width=0.45\textwidth]{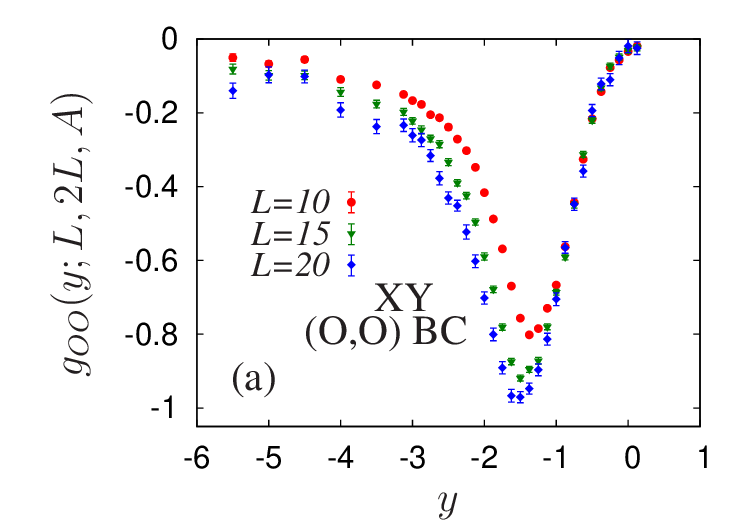}
\includegraphics[height=8cm,width=0.45\textwidth]{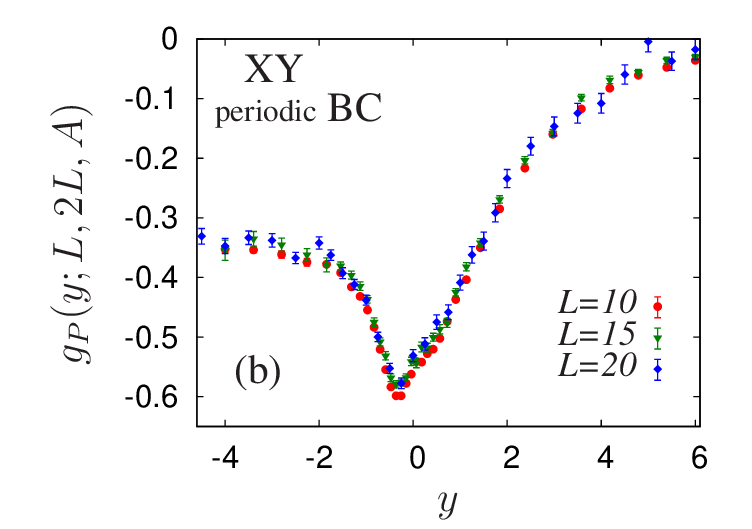}

\caption{Monte Carlo data for
 $g(y=\tau(L-\frac{1}{2})^{1/\nu};L,2L,A=(L/\rho)^2)$ (see Eq.~\reff{eq:defg},
 $\tau=(T-T_c)/T_c$) in the three-dimensional XY model for $L=10$, $15$, $20$,
 and fixed inverse aspect ratio $\rho^{-1}=6$. In (a) and (b) we present the
 result for $(O,O)$ and  periodic BC, respectively.  For $T\ge T_c$, i.e.,
 $y=y_+>0$ one has $y_{+}=[(L-\frac{1}{2})\xi^+_0/\xi_+]^{1/\nu}$. For the XY
 model $\xi_-=\infty$ for all temperatures $T\le T_c$. 
 The Kosterlitz-Thouless transition of the two-dimensional film occurs at
 $y=y_{c,OO} = -2.69(3)$~\cite{Hasen} and $y=y_{c,P}= -0.996(1)$~\cite{SM-95} 
 in (a) and (b), respectively.
}
\label{fig:xy_gy}
\end{figure}

\begin{figure}
\includegraphics[height=7cm,width=0.45\textwidth]{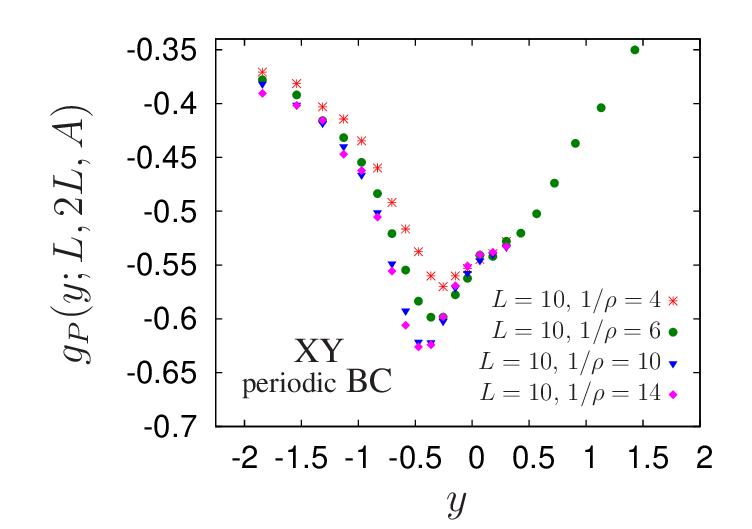}
\caption{Monte Carlo data for
  $g(y=\tau(L-\frac{1}{2})^{1/\nu};L,2L,A=(L/\rho)^2)$ (see
  Eq.~\reff{eq:defg}, $\tau=(T-T_c)/T_c$) within the three-dimensional XY
  model with periodic BC for $L=10$ and different values of the  inverse
  aspect ratio $\rho^{-1}$. For $y\ge 0$ one has
  $y=y_+=[(L-\frac{1}{2})\xi_0^+/\xi_+]^{1/\nu}$. For the XY model
  $\xi_-=\infty$ for all temperatures $T\le T_c$. 
 The Kosterlitz-Thouless transition of the two-dimensional film occurs at
 $y=y_{c,P}= -0.996(1)$~\cite{SM-95}.
 Note the enlarged scales as compared  with Fig.~\ref{fig:xy_gy}(b).}
\label{fig:xy_a}
\end{figure}

%
In Fig.~\ref{fig:xy_gy}  we report the data  corresponding to 
$g(y;L,2L,A)$ (see Eq.~\reff{eq:defg}) for (a) $(O,O)$ and (b) periodic BC.
Corrections to scaling, which are signaled by the fact that data corresponding
to different $L$ do not fall onto the same master curve, are much more
pronounced for  the case of $(O,O)$ BC (see Fig.~\ref{fig:xy_gy}(a)) as 
compared with 
the case of periodic BC  (see Fig.~\ref{fig:xy_gy}(b)). The same holds
for the dependence of the data on the aspect ratio $\rho$ 
(data for $(O,O)$ are not shown, 
data for periodic BC are presented in Fig.~\ref{fig:xy_a}).
For $(O,O)$ and periodic BC corrections to scaling are  more relevant
 for $y\lesssim y_{\rm min}$, where $y_{\rm min}$ is the value of $y$ at which 
 the function  $g(y;L,2L,A)$ attains its  minimum. 
For $y\gtrsim y_{\rm min}$ the data obtained for different $L$ follow a
common curve. We note that the bulk correlation length $\xi$ 
of the XY model is
infinite for all temperatures below $T_c$: $\xi(T\le T_c) =
\infty$. Accordingly, within the XY model the scaling variable $y$ can be
expressed as $y= [(L -1/2)\xi_0^+/\xi]^{1/\nu}$ only for $T\ge T_c$.

Interestingly, for both types of BC  the aspect ratio dependence is 
particularly strong in  the range
of temperatures around the minimum of the function $g(y;L,2L,A)$, i.e., 
$-2\lesssim y\lesssim -1$ and $-1\lesssim y\lesssim 0$ for 
$(O,O)$ and periodic BC, respectively   (see Fig.~\ref{fig:xy_a}).
According to Fig.~\ref{fig:xy_gy}, the minimum of $g(y;L,2L,A)$ for periodic BC
occurs at the reduced temperature
 $\tau_{\rm min} =-y_{\rm min}/(L-\frac{1}{2})^{1/\nu}\simeq -0.31/(L-\frac{1}{2})^{1/\nu}$.
For $(O,O)$ BC it occurs slightly further away
from $T_c$, i.e., at  $y_{\rm min}\simeq -1.34$ and $ -1.50(3)$, 
depending on the value of $L$.  
We  find   that  changing $L$
at a fixed aspect ratio results in   slight relative shifts 
of the data  whereas changing $\rho$ at fixed
$L$ leads to   much more pronounced differences
(see Fig.~\ref{fig:xy_a}).
This behavior is expected to be related to
the finite-size effects near  the thin {\it film} critical point.
Within the Ising model, for an infinitely large transverse area $A$
the point at which the film  with $(O,O)$ or  periodic BC exhibits the 
2D critical behavior is located on the bulk coexistence line $H=0$ at a
size-dependent temperature $T_c(L) < T_c$ such that $\xi(T=T_c(L))\sim
L$. Accordingly, upon increasing $L$, $T_c(L)$ approaches the
three-dimensional bulk value $T_c$ as $T_c(L\rightarrow\infty) = T_c (1 + y_c
L^{-1/\nu})$ \cite{FSS}, where $y_c$ is negative, {\it non-universal}, and
depends, inter alia, on the type of BC. 
The corresponding scaling variable
$x(L)\equiv [(T_c(L) - T_c)/T_c] (L/\xi_0^+)^{1/\nu}$ tends to a {\it
universal} and BC-dependent value $x^*\equiv x(L\rightarrow\infty) = y_c
(\xi_0^+)^{-1/\nu}$.
Hence, $y_c$  is expected to lie 
in the vicinity of  the minimum  of the function  $g(y;L,2L,A)$.
Accordingly, around its  minimum the function  $g(y;L,2L,A)$ should
exhibit a strong dependence  on  the aspect ratio $\rho$ if
the bulk correlation length $\xi^\mathrm{2D}$,  associated with the shifted
critical point of the two-dimensional film \cite{ES}, becomes comparable with
the characteristic  transverse length $L_{\parallel}\equiv \sqrt A$ of the
simulated system.
%

Within the XY model  the critical point of the thin film  belongs to 
the Kosterlitz-Thouless (KT) universality class \cite{K-T}.
The KT theory predicts that upon approaching this critical point from the
high temperature phase the correlation length $\xi_{KT}\sim \exp
[(1-\beta/\beta_c^{KT})^{-\nu^{KT}}]$, $\nu^{KT}=1/2$, diverges
exponentially. 
The shift of $T_c(L)$ relative to the bulk critical point is expected to
scale with the film thickness $L$ in  the same way as for the Ising model,
i.e., $(T_c(L) - T_c)/T_c \simeq  y_c L^{-1/\nu}$ for large $L$, with
$\nu=0.662(7)$ for the 3D XY model. 
This prediction is in agreement with  MC simulations of various models
belonging to the XY universality class and confined in films
with free~\cite{janke,SS-92,Hasen} or periodic~\cite{SM-95} BC. 
The critical exponent $\nu$ obtained from early simulations~\cite{janke,SS-92}
of films with free BC was slightly larger ($\nu \simeq
0.7$~\cite{janke,SS-92}) than the theoretically predicted value
$\nu=0.662(7)$, due to rather strong corrections to scaling which need to be
taken into account in order to observe the theoretically expected
behavior~\cite{Hasen}. The results of Ref.~\cite{Hasen} for
$x^*_{OO}=-7.64(15)$ yield the estimate $y_{c,OO}=-(\xi_0^+)^{1/\nu}x^*_{OO} =
-2.69(3)$ for the location of the shifted KT transition in
the XY model with free boundary conditions, whereas $y_{c,P}=-0.996(1)$ for
PBC~\cite{SM-95}. 
It turns out that in the simulations with free BC~\cite{janke} 
the positions of the 
maxima of the  thermodynamic  functions such as the peak of the specific heat
or the peak of the susceptibility do not coincide with the transition point
but occur at ca. $1.3 \times T_c(L)$. (These quantities are not related to
singularities of XY films.)  
With  increasing film thickness $L$ the absolute distance in temperature 
of these peaks  from $T_c(L)$ decreases and for $L=10$ the simulations
of Ref.~\cite{janke} report a shift of  less than 10$\%$. There is also
experimental evidence that as a  function of temperature the position
of the minimum of the Casimir force of $^4$He films, which belong to the
universality class of XY films, coincides with the position $T_m(L)$ of their
specific heat maximum (see Subsec.~VD and Figs. 21 and 32 in
Ref.~\cite{gasparini}),  whereas the onset of superfluidity in these films
occurs at $T_c(L)<T_m(L)$ (see Subsec.~VD and Figs. 24, 32, and 33 in
Ref.~\cite{gasparini}).
A similar behavior may be expected to hold for the function
$g(y;L,2L,A)$. Indeed,  as can be seen from Figs.~\ref{fig:xy_gy} and 
\ref{fig:xy_a}, the minima of
the function $g(y;L,2L,A)$ lie in the vicinity of the corresponding values of
$y_c$. Therefore, similar to the Ising model,  the strong aspect ratio
dependence around the minimum might occur when the exponentially diverging
bulk correlation length $\xi^\mathrm{2D}$,  associated with the KT critical
point of the film, becomes comparable with the characteristic  transverse
length $L_{\parallel}\equiv \sqrt A$ of the  simulated system. 

\subsubsection{Dirichlet-Dirichlet boundary conditions}

\begin{figure}
\includegraphics[height=7cm,width=0.45\textwidth]{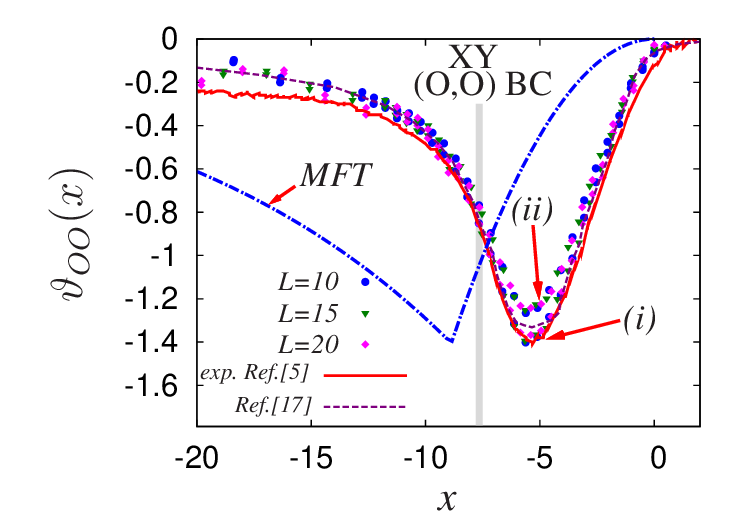}
\caption{Scaling function $\sf_{OO}$ of the Casimir force for the
three-dimensional XY model with $(O,O)$ BC. The MC data reported in this figure refer to lattices with
$L=10$, $15$, and $20$, with fixed inverse aspect ratio $1/\rho=6$.
Corrections to scaling have been accounted for according to two different
ans{\"a}tze, provided by Eq.~\reff{eq:xy_f_fit1} and  Eq.~\reff{eq:xy_f_fit2};
the corresponding numerical results are denoted by (i) and (ii), 
respectively.
With corrections to scaling  of the form (ii), the {\it shape} of the
 resulting scaling function is almost indistinguishable from the one obtained
 with corrections to scaling  of the form (i), but its overall amplitude  is
 reduced by a factor $R\simeq 0.9$.
For (i) our MC data compare very well with the corresponding experimental data
from Ref.~\cite{garcia} (solid line) and with the MC data of Ref.~\cite{hucht}
(dashed). Due to the Goldstone  modes 
$\sf_{OO}(x\to-\infty)=\mathrm{const}\ne 0$.
The dash-dotted 
line shows the mean field   scaling function \cite{LGW-MF,LGW-MF1}
normalized to  the depth of the minimum of the MC data (i). The levelling off
of the experimental data \cite{garcia} for $x\to -\infty$ contains a component
which is specific for $^4$He wetting films \cite{kardar:04} and cannot be
captured by an XY lattice model. 
The gray bar indicates the position and uncertainty of the universal 
value $x^*_{O,O} = -7.64(15)$ 
of the scaling variable $x$ corresponding to the occurrence of the
Kosterlitz-Thouless transition in the film, 
as inferred from MC simulations of
lattice models in the XY universality class presented in Ref.~\cite{Hasen}. 
}
\label{fig:xy_scaling}
\end{figure}

We consider first the case of $(O,O)$ BC.
As evidenced 
by Fig.~\ref{fig:xy_gy}(a), in order to achieve a good data collapse
of the curves  corresponding to  different lattice sizes we have to account
for corrections to  scaling  according to Eqs.~\reff{eq:xy_x_fit}
and~\reff{eq:xy_f_fit}. As a phenomenological ansatz for the effective
corrections we take $\omegaeff=1$   and consider two
functional forms for the $L$-dependent corrections to the scaling function:
case (i) [Eq.~\reff{eq:xy_f_fit1}] and case (ii) [Eq.~\reff{eq:xy_f_fit2}]
as discussed in Subsec.~\ref{subsec:corr}.
As a result of the fitting procedure, in the interval 
$x\in [-6,-2.1]$ (see Eq.~(\ref{eq:xy_x_fit}))
 we find $r_1=1.18(10)$, $r_2=2.40(13)$,
$g_{1}=5.83(25)$, and $g_{\omega}=2.25(15)$ in case (i) and
$g_{2}=-2.98(8)$ in case (ii) with the same values for $r_1$, $r_2$, and
$g_\omega$ as in case (i). 
Figure~\ref{fig:xy_scaling} shows  the corresponding 
resulting estimates of the scaling function  $\sf(x)$ of the critical 
Casimir force.  
The quality of the data collapse for the two cases separately clearly
indicates that Eqs.~\reff{eq:xy_f_fit1}, \reff{eq:xy_f_fit2},
and~\reff{eq:xy_x_fit} are very effective ways of accounting for the
corrections to scaling in this system.
We find that  $\sf(x)$ is  slightly affected by the choice of the functional
form of corrections to scaling and indeed in the two cases one finds estimates
of $\sf(x)$ which have the same shape but the overall amplitude  is reduced by
a factor $R\simeq 0.9$ in case (ii) as compared with case (i).
The dashed line represents the scaling function which has been
determined in Ref.~\cite{hucht} on the basis of a different numerical method
and assuming corrections to scaling of the form (i). 
Even though this result is actually biased by that particular choice (a point
which has not been discussed in Ref.~\cite{hucht}),  
the very good agreement between the different approaches
provides a highly welcome independent test of both methods.
Our MC results for $\vartheta(x)$ compare well also with the experimental
data of Ref.~\cite{garcia}.
(For a meaningful comparison between the numerical and the experimental
scaling function, the abscissa $\tau L^{1/\nu}$ of the experimental data
presented in Ref.~\cite{garcia} has to be properly normalized as
$\tau(L/\xi_0^{+\mathrm{(exp)}})^{1/\nu}$ by using the experimental value
$\xi_0^{+\mathrm{(exp)}} = 1.432$\AA~\cite{ahlers,footnote}.)
In particular, the position of the pronounced minimum of the scaling function
is properly captured.
The corrections to scaling of form (i) yield
 $x^{(i)}_\mathrm{min} = -5.43(2)$
and $\sf^{(i)}_\mathrm{min} \equiv \sf(x^{(i)}_{\mathrm{min}}) = -1.396(6)$, 
whereas those of form (ii) result in $x^{(ii)}_\mathrm{min} = -5.43(2)$
and $\sf^{(ii)}_\mathrm{min} \equiv \sf(x^{(ii)}_{\mathrm{min}}) = -1.260(5)$. 
The corresponding experimental values are
$x_{\mathrm{min}}^\mathrm{(exp)} = -5.7(5)$ and
$\sf_\mathrm{min}^\mathrm{(exp)} = -1.30(3)$. 
Taking into account the aforementioned 
 bias affecting the results of Ref.~\cite{hucht} and
the sensitivity of the resulting scaling function to the assumed
form of the corrections to scaling we conclude that our estimates for
$x_\mathrm{min}$ and $\sf_\mathrm{min}$ are compatible also with 
those presented there ($-5.3(1)$ and $-1.35(3)$, respectively).
As expected, due to the presence of the Goldstone modes  below $T_c$,
both the experimental and the MC data do not approach zero for
$x\rightarrow-\infty$ 
but saturate at some finite negative value at  low temperatures.
However, the absolute value of the saturation  as obtained 
from the MC simulations is
 smaller than the experimental one. This difference, which extends deep into
 the non-critical regime, is, inter alia, due to $^4$He specific properties
 and to  the occurrence of capillary waves on the liquid-vapor interface of
 the critical $^4$He wetting films. This point has been discussed  in
 Ref.~\cite{kardar:04}.
In Fig.~\ref{fig:xy_scaling} the gray vertical bar indicates the {\it
universal}  value $x^*_{O,O} = -7.64(15)$ of the scaling variable
corresponding to the occurrence of the Kosterlitz-Thouless transition at
$T=T_c(L)$ in the film, as inferred from MC simulations of lattice models
in the XY universality class presented in Ref.~\cite{Hasen}.
The Kosterlitz-Thouless transition is accompanied by an actually 
invisible essential 
singularity $\sim \exp(-{\rm const}/\sqrt{|x-x^*|})$ in the behavior
of the specific heat which, as discussed above,
displays a pronounced maximum at a temperature $T=T_m(L)$. 
Accordingly, one does
not expect to find any particular signature of this transition in the scaling
function of the Casimir force for $x\simeq x^*$, in distinction to the case 
of the Ising model (c.f., Subsecs.~\ref{subsec:DD} and~\ref{subsec:IsP}).

Finally, for  completeness,
in  Fig.~\ref{fig:xy_scaling} we have  also included (dash-dotted line) our
mean field  result for the Casimir 
scaling function $\vartheta_{OO}^{\mathrm{(MFT)}}$ obtained from the limiting
case of
the  vectoralized Blume-Emery-Griffiths lattice model corresponding
to the model of  pure $^4$He \cite{LGW-MF}.
The scaling function is normalized to the depth of the minimum of the MC data.
 For large $L$,  $\vartheta_{OO}^{\mathrm{(MFT)}}$ 
agrees very well with the ones obtained from  the
$O(2)$ Landau-Ginzburg continuum theory \cite{LGW-MF,LGW-MF1}.

\subsubsection{Periodic boundary conditions}

\begin{figure}
\centering
\includegraphics[height=7cm,width=0.45\textwidth]{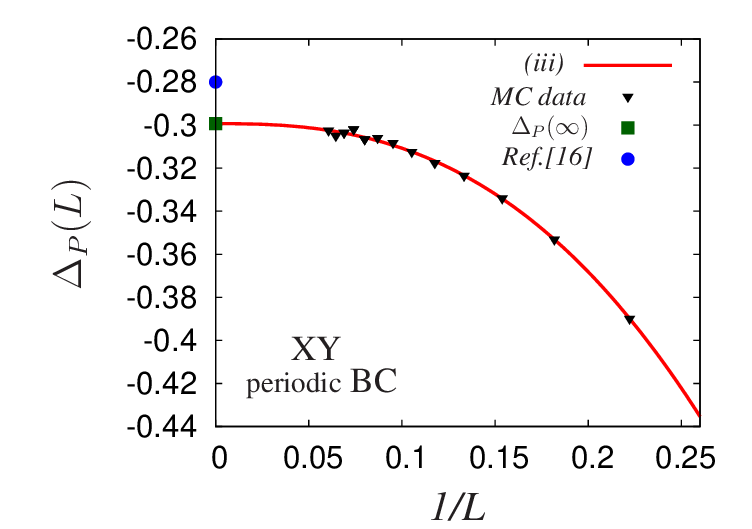}
\caption{%
Critical Casimir amplitude $\Delta_P(L)$ for the three-dimensional XY
  bulk universality class and periodic BC, estimated from
  lattices of several thicknesses $L$  and inverse aspect ratio $1/\rho=6$.
Due to  corrections to scaling, $\Delta_P$ depends on $L$. 
The solid line represent the best fit to the numerical data based on
Eq.~\reff{eq:fit_delta_new} and allows one to extrapolate the value of
$\Delta_P(L)$ to the scaling limit $L\rightarrow\infty$, resulting in $\Delta_P
= -0.2993(7)$ ($\blacksquare$).
With $\bullet$ we indicate the numerical estimate  $\Delta_P =-0.28$
provided in Ref.~\protect{\cite{DK}}. 
}
\label{fig:delta_fit_per}
\end{figure}

In this subsection we discuss the XY model with  periodic BC.
According to Fig.~\ref{fig:xy_gy}(b)
corrections to scaling are much less
pronounced in this case than for  $(O,O)$ BC (Fig.~\ref{fig:xy_gy}(a)), 
suggesting that the exponent $\omegaeff$ might be actually larger than $1$.
In addition, the dependence of the numerical data on the
aspect ratio $\rho$ turns out to be relevant only in the restricted
range $-1\lesssim y \lesssim 0$  of the scaling variable 
(see Fig.~\ref{fig:xy_a}), so that   the assumed forms of the aspect ratio
corrections
 in Eqs.~\reff{eq:xy_x_fit} and~\reff{eq:xy_f_fit} do not work best.
In the present case, the accuracy of our Monte Carlo data 
allows us to study in some detail also the Casimir amplitude $\Delta\equiv
\sf(0)/2$. 
Upon focusing on such a quantity in a broader range of geometries 
$(6\le L\le 20)$ 
it turns out that for this amplitude the 
corrections to scaling are not properly accounted for 
by the previous ans\"atze (case (i) and case (ii), Eqs.~\reff{eq:xy_f_fit1}
and  \reff{eq:xy_f_fit2}, respectively). 
We have therefore tried also a fit of the exponent
$\omegaeff$ according to Eq.~\reff{eq:xy_f_fit} with  $r_{1,2}=0$ (case (iii),
Eq.~\reff{eq:xy_f_fit3}),
which yields for the Casimir amplitude
\be
\label{eq:fit_delta_new}
\Delta(L)=\Delta (1+ g_3 L^{-\omegaeff}).
\ee
With this ansatz, our data for $\Delta_P(L)$ are very well fitted for
$\omegaeff = 2.59(4)$ and 
$g_3=14.9(7)$ in the interval $0\le L^{-1} \le 0.15$.
(At present, the origin  of
this rather large value of $\omegaeff$ is not clear.)
The comparison between the numerical data and the fit is reported 
in Fig.~\ref{fig:delta_fit_per}.
The value of the Casimir amplitude extrapolated to the scaling limit
$L\rightarrow\infty$ is $\Delta_P(\infty)\equiv \Delta_P = -0.2993(7) $ which is slightly smaller
than the previous estimate $\Delta_P = -0.28$ 
(see Ref.~\cite{DK} and the discussion below). 
Note, however, that our estimate is biased by
the particular form Eq.~\reff{eq:fit_delta_new} 
assumed for the corrections to scaling. 

\begin{figure}
\includegraphics[height=7cm,width=0.45\textwidth]{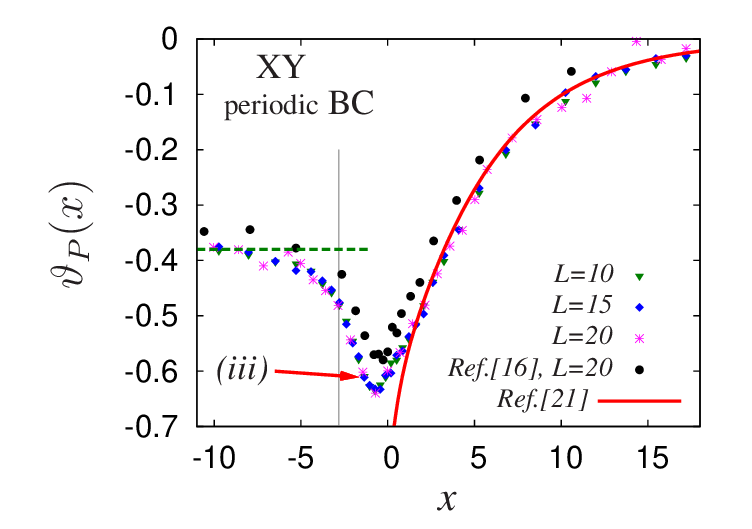}
\caption{Scaling function $\sf_{P}$ of the Casimir force for the
three-dimensional XY model with periodic BC.
The corrections to scaling are taken into account by Eq.~\reff{eq:xy_f_fit3} (case (iii)) and Eq.~~\reff{eq:xy_x_fit} with $r_1=0$. 
The shape of our MC data compares very well with the corresponding  MC data
($\bullet$) of Ref.~\cite{DK}. For a discussion of the relative shift of the
data sets see the main text.
The solid line corresponds to the analytical prediction in Ref.~\cite{GD}.
Due to the Goldstone modes, in agreement with Eq.~\reff{eq:XYperplat}, 
$\sf_P(x\to-\infty)= -0.383(4)$, see horizontal dashed line. 
Contrary to $(O,O)$ BC in Fig.~\ref{fig:xy_scaling}, for periodic BC MFT yields
$\sf_{P}^{\mathrm{(MFT)}}(x)\equiv 0$ for $x\lessgtr 0$.
The gray vertical line indicates the position of the universal value 
$x^*_P = -2.82(2)$ of the scaling variable $x$
corresponding to the occurrence of the Kosterlitz-Thouless transition
in the film, as inferred from MC simulations~\cite{SM-95}.
}
\label{fig:xy_scaling_per}
\end{figure}

The analysis of the Casimir amplitude $\Delta_P(L)$ suggests that the
corrections to scaling for periodic BC are well captured (in the range of
sizes and of the  scaling variable investigated here) by
Eq.~\reff{eq:xy_f_fit3} (case (iii))
and  Eq.~\reff{eq:xy_x_fit} with $r_{1}=0$. 
The resulting estimate for the scaling function $\sf_P$ is
reported in Fig.~\ref{fig:xy_scaling_per} for which
 we adopt the values for 
$g_3$ and $\omegaeff$ which we determined from the analysis of the
correction to scaling for $\Delta_P(L)$.
It turns out that a very good data collapse is achieved even
without correcting the abscissa, i.e., with $g_\omega\simeq 0$, $r_1\simeq 0$,
within the range of the scaling variable $x$ we have investigated,
which actually includes the interval $-1<y<0$ in which
 the corresponding function
$g$ shows a more pronounced dependence on $\rho$.

As another valuable test of the method, our results are compared
with the  corresponding MC simulation data obtained previously in
Ref.~\cite{DK} within a
different approach, i.e., by computing the average value of the lattice
stress-tensor.  In Fig.~\ref{fig:xy_scaling_per} we report the data set 
corresponding to the  lattice size $L=20$  investigated therein.
The shapes of the two scaling functions are 
very similar but the data points from
Ref.~\cite{DK} are shifted upwards with respect to the ones we have 
obtained.
This discrepancy might be due to the uncertainty in the 
normalization factor used in Ref.~\cite{DK}, where
the vertical scale of the data for $\vartheta_P$ has to be adjusted 
on the basis of an independent estimate. This estimate has been
obtained  from the 
$\epsilon=4-d$-expansion of the ratio  $\Delta_{P,n}/\Delta_{P,1}$
of the Casimir amplitudes for $O(n)$ models with the result
$\Delta_{P,n=2} = -0.28$ so  that
$\sf_{P}(0)\equiv 2\Delta_{P,n=2}=-0.56$. 
In contrast, the method presented here provides absolute values of the amplitude and the scaling function.
In addition to the uncertainty concerning 
the normalization factor, in Ref.~\cite{DK}
no corrections to scaling have been applied in the
determination of $\sf_P$.
The present MC results provide the
estimates
$x_\mathrm{min}=-0.73(1)$ and  $\sf_{P}(x_\mathrm{min}) =-0.633(1)$ 
characterizing 
the position of the minimum of the scaling function.

For the scaling function of the XY model with periodic BC 
some analytical predictions are also available; for a thorough 
comparison of the scaling function obtained within various approaches
see  Ref.~\cite{DKD}. Here we discuss only the comparison with the recent
results based on a suitable perturbation theory for the $O(n)$ model in a film
geometry with periodic BC~\cite{DGS,GD}, which improves previous
analyses~\cite{krech:92,KD-92} of
this scaling function for  $T\ge T_c$ by taking into account a higher-order
contribution to the perturbation theory which involves fractional powers of
$\epsilon$.
In the case $n=2$ (XY model) and in agreement with our MC data this latter
analytically available scaling function decreases
 monotonically for $x \rightarrow
0$ and thus allows for the formation of a minimum below $T_c$ (without being
able to reach it) whereas the previously available analytic scaling function
exhibits a minimum above $T_c$.
The analytically estimated value for the critical Casimir amplitude is 
$\Delta_{P} \simeq - 0.43$ (i.e.,  $\sf_{P}(0) \simeq -0.86$) 
which in absolute value is larger than the MC result. 
In Fig.~\ref{fig:xy_scaling_per} this  analytically
predicted scaling function is reported, for comparison, as a solid line.

As already mentioned above, one characteristic feature of the scaling
function of the critical Casimir force in
the XY model (and, more generally, in systems with continuous symmetry) 
is its saturation at a nonzero 
negative value $\sf(x\rightarrow \infty) < 0$ at low
temperatures, which occurs for all non-symmetry breaking BC.
This is due to the fact that, even well below the critical temperature, the
fluctuations of the order parameter exhibit long-ranged 
correlations due to the Goldstone modes associated with the broken continuous 
symmetry, which result in a non-vanishing long-ranged Casimir force.
For periodic BC the saturation value $\sf_P(-\infty)$ 
is significantly more negative 
and is approached more rapidly than in the case of $(O,O)$ BC.
The line of arguments presented in Ref.~\cite{kardar:04} for the
theoretical calculation (TH) of 
$\sf^\mathrm{(TH)}_{O,O}(-\infty) =  - \zeta(3)/(8\pi) \simeq -0.049$ 
(disregarding additional
helium-specific surface
fluctuations) can be extended to the present case by considering periodic
(instead of Neumann as in Ref.~\cite{kardar:04}) 
BC for the fluctuations of the {\it phase} field of the
order parameter in the film. 
In three dimensions this yields 
\be
\sf^\mathrm{(TH)}_P(-\infty) = 2
\Delta^\mathrm{(G)}_P \simeq -0.38, 
\label{eq:XYperplat}
\ee
where
$\Delta^\mathrm{(G)}_P = - \zeta(3)/(2\pi) \simeq -0.19$ 
is the Casimir amplitude for a one-component ($N=1$) fluctuating Gaussian
field in a film with PBC (see, e.g., Eq.~(9.2) in Ref.~\cite{KD-92}), so that
$\sf^\mathrm{(TH)}_P(-\infty)/\sf^\mathrm{(TH)}_{O,O}(-\infty) = 8$.
The numerical data corresponding to the MC simulations presented in
Fig.~\ref{fig:xy_scaling_per} yield $\sf_P(-\infty) = -0.383(4)$ 
(obtained by fitting the data points in the region $-14<x\le
-10$, two of which are actually not shown in Fig.~\ref{fig:xy_scaling_per}, 
with a constant). 
This is in very good agreement with the theoretical prediction
$\sf^\mathrm{(TH)}_P(-\infty)$ in Eq.~\reff{eq:XYperplat}. 
Note that the MC data of Ref.~\cite{DK} give
$\sf_P(-\infty) \simeq -0.33$, a value which is biased by the choice of the
normalization of the scaling function, as mentioned before.
We point out, however, that the line of arguments in Ref.~\cite{kardar:04}
assumes that, deep in the low-temperature phase, the phase field obeys Neumann
BC and that the magnitude of the complex
order parameter (superfluid density) is spatially constant across the 
film, i.e.,
that the effects of the surfaces are effectively negligible. This might not be
the case in the presence of the Goldstone modes which can cause the magnitude
of the order parameter vary algebraically within the film.
%
%
%
%

Finally, in Fig.~\ref{fig:xy_scaling_per} 
we report as a gray vertical line the {\it
universal}  value $x^*_P = -2.82(2)$ of the scaling variable $x$
corresponding to the occurrence of the Kosterlitz-Thouless transition in the
film, as inferred from the MC simulations of the XY model 
in a film with PBC~\cite{SM-95}. As in the case of $(O,O)$ BC, there is no
singularity possibly visible 
in $\sf_P(x)$ associated with this transition.  

\subsection{Ising model}
\label{subsec:ising}
In the case of the Ising model we have determined the scaling function $\sf$
for $(+-)$, $(++)$, Dirichlet-Dirichlet $(O,O)$, and periodic BC.
The first two BC are relevant for interpreting the results of the
experiments in Refs.~\cite{Hertlein,pershan} 
which use as a critical medium classical binary liquid mixtures
near their demixing point.

In our simulations we have used lattices with
$L=10$, $13$, $16$, and $20$ and with  $L_x=L_y=6L$, i.e., $\rho=1/6$.  
Each data point has been averaged over at least $10^5$ hybrid MC steps.
\begin{figure}
\includegraphics[height=7cm,width=0.45\textwidth]{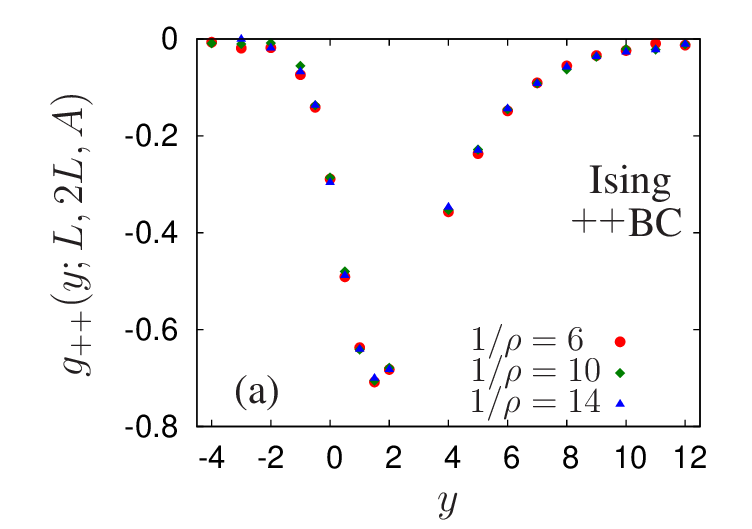}

\includegraphics[height=7cm,width=0.45\textwidth]{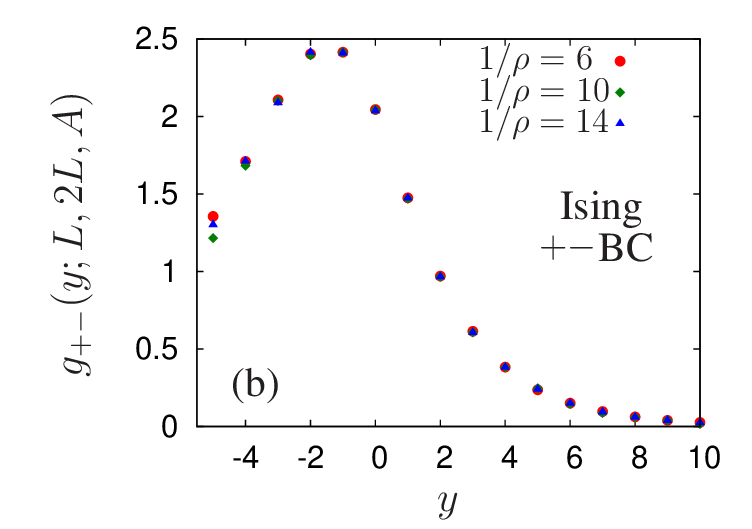}
\caption{Plot of  $g(y;L,2L,A =(L/\rho)^2)$
(see Eq.~\reff{eq:defg}) 
for the three-dimensional Ising model with  $L=10$ and 
$1/\rho = 6$, $10$, and
$14$.  (a) and (b) refer to $(++)$ and $(+-)$ BC, respectively, and the
coincidence of data points corresponding to different values of $\rho$  
demonstrates that the geometry of the lattice does not affect the resulting
finite-size critical behavior in the 
region $-4\lesssim y \lesssim 10$. 
}
\label{fig:ising_asp_pp}
\end{figure}

\subsubsection{$(++)$ and $(+-)$ boundary conditions}

%
%
We first discuss the cases of $(++)$ and $(+-)$ BC,
for which  we find that in the  critical regime 
the numerical data for the function $g(y;L,2L,A)$ are practically independent 
of the aspect ratio $\rho = L/\sqrt{A}$  
(see Fig.~\ref{fig:ising_asp_pp}). 
The  presented data correspond to $L=10$  and to inverse 
 aspect ratios $\rho^{-1}=6,10,14$. 
In the case of $(+-)$ BC  the aspect ratio becomes
relevant for $y\lesssim -4$, where the behavior of
the system is dominated by the presence of the strongly fluctuating 
interface which separates the regions with predominantly positive and negative
magnetization. The extent of these fluctuations is known to be particularly
sensitive to the spatial extension and to the 
 geometry of the system in the directions
parallel to the interface (i.e., in the $L_x$ and $L_y$ directions); therefore the
aspect ratio $\rho$ plays an important role for these fluctuations. 
In $d=3$ one expects a strongly increasing parallel
correlation length $\xi_{\parallel}$ 
which  governs the decay of the correlations in the direction
parallel to the interface, 
i.e., $\xi_{\parallel} \sim \exp (L_{\parallel}/(4\xi))$ with $L_{\parallel} = L_x = L_y$~\cite{parryevans}.
In addition, these strong interfacial fluctuations 
cause the scaling function $\sf_{+-}$ to decay to zero for
$x\rightarrow -\infty$ much more 
slowly than the scaling function $\sf_{++}$ (see, c.f.,
Figs.~\ref{fig:ising_pm} and~\ref{fig:ising_pp}).

Contrary to the aspect ratio, $L$-dependent 
corrections to scaling are rather important
for the Ising model with $(++)$ and $(+-)$ BC. 
By using the phenomenological  
ans\"atze in Eqs.~\reff{eq:xy_x_fit} and~\reff{eq:xy_f_fit1} 
or~\reff{eq:xy_f_fit2} with $r_{1,2} = 0$ (which account for the negligible
dependence of the data on $\rho$) 
we have obtained a good data collapse for the scaling functions calculated
for $L= 13$, $16$, and $20$.
However, these ans\"atze fail to describe the data for the critical Casimir
amplitude $\Delta$ in the broader range of 
thicknesses $6 \le L \le 20$, as it is the case of the XY model with periodic
BC.
%
\begin{figure}
\includegraphics[height=7cm,width=0.45\textwidth]{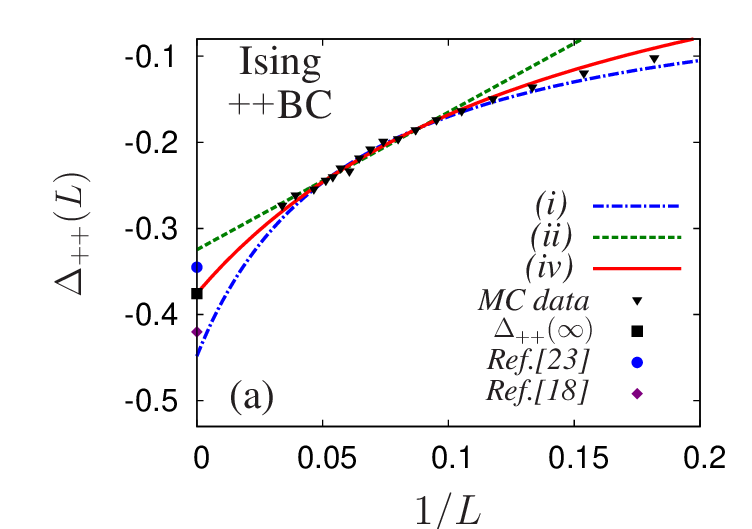}

\includegraphics[height=7cm,width=0.45\textwidth]{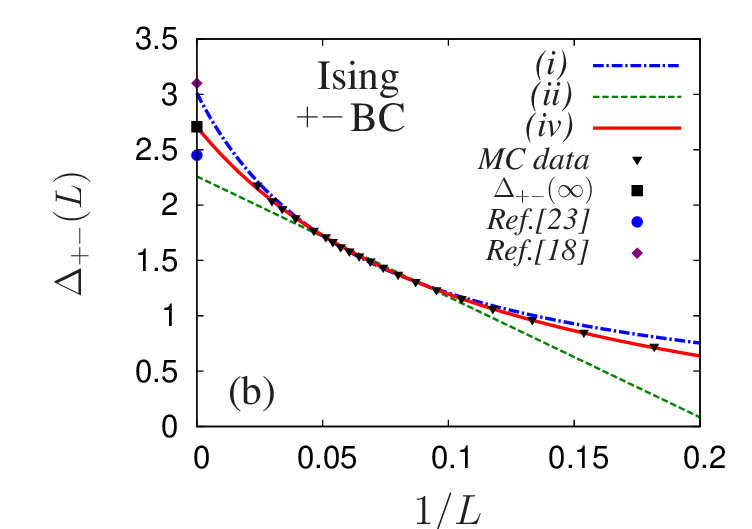}
\caption{%
MC data for the critical Casimir amplitude 
$\Delta(L)$ of the three-dimensional Ising model with (a) $(++)$ and (b) $(+-)$
BC, as a function of the inverse lattice size $L$ (for lattices with fixed inverse 
aspect ratio $1/\rho =6$).
 $L$-dependent corrections to scaling give rise to the dependence  $\Delta(L)$ such that 
$\Delta_{++/+-} \equiv \Delta_{++/+-}(L\to \infty)$.
The solid line corresponds to the best fit obtained by using the fitting
ansatz in Eq.~\reff{eq:fit_delta_new_r} in the interval $0<1/L\le 0.1$. For
comparison we present also the best fits using  the ans{\"a}tze (i) $\Delta(L)=\Delta
(1+g_1L^{-1})^{-1}$ and  (ii) $\Delta(L)=\Delta (1+g_2L^{-1})$. 
Our estimates ($\blacksquare$) for the asymptotic values of the Casimir
amplitudes compare reasonably well with previous MC results ($\bullet$) from
Ref.~\protect{\cite{krech}} and with results  ($\blacklozenge$) obtained 
from the 
de Gennes-Fisher local-functional \protect{\cite{upton}}. 
}
\label{fig:delta_fit}
\end{figure}
%
It turns out that the corrections to scaling in this range are very well
captured by the functional dependence (iv) (Eq.~\reff{eq:fit_new}
introduced in
Subsec.~\ref{subsec:corr}, with $r_{1,2}=0$)
which for the  critical Casimir amplitude yields
\be
\label{eq:fit_delta_new_r}
\Delta(L)=\Delta \frac{(1 + \tilde g_{1}L^{-1})}{(1+{\tilde g}_{2}L^{-1})}.
\ee
As  in the case of the XY model with periodic BC we shall determine
the parameters $\tilde g_1$ and $\tilde g_2$
 (according to Subsec.~\ref{subsec:corr}) for both $(++)$ and
$(+-)$ BC
on the basis of the analysis of the
corrections to scaling to the corresponding Casimir amplitude and then these
values are employed in order to calculate the scaling functions
$\sf_{++}$ and $\sf_{+-}$.  

In Figure~\ref{fig:delta_fit} we present numerical data for $\Delta$ 
as a function of $1/L$ for both $(++)$ (a) and $(+-)$ (b) BC with the
corresponding fit carried out according to Eq.~\reff{eq:fit_delta_new_r}
in the interval $0\le 1/L \le 0.1$.
Other variants of the fit function (within the same fit interval), such as
$\Delta(1+ g_{1}L^{-1})^{-1}$
and $\Delta(1+ g_{2}L^{-1})$, indicated as (i) and (ii),
respectively, are also presented for comparison. 

For $(++)$ BC  the fitting parameters are 
$\tilde g_1 = -2.6(1.2)$,
$\tilde g_2 = 6.6(3.7)$ and
the resulting estimate for the Casimir amplitude is
$\Delta_{++}(L\to\infty)\equiv \Delta_{++} =  -0.376(29)$, i.e., $\sf_{++}(0) = -0.75(6)$, which
compares  quite well with the previous MC
result  $\sf_{++}(0) = -0.690(32)$~\cite{krech} 
shown as a full circle in
Fig.~\ref{fig:delta_fit}(a); 
for the latter result corrections to scaling were not taken into account.
Field-theoretical
predictions $\sf_{++}^\mathrm{(FT)}(0) = -0.652\ldots -0.346$  give
numbers slightly smaller in absolute value which depend on
the approximant used to re-sum the  field-theoretical $\epsilon
= 4-d$-expansion up to $O(\epsilon)$ series 
(see Ref.~\cite{krech} for details).

For $(+-)$ BC we have found $\tilde g_1 = -1.8(1)$,  $\tilde g_2 = 8.54(43)$
and we estimate $\Delta_{+-}(L\to\infty)\equiv\Delta_{+-} = 2.71(2)$, i.e., $\sf_{+-}(0) = 5.42(4)$, in
agreement with the experimental value $\sf^\mathrm{(exp)}_{+-}(0)
=6(2)$~\cite{pershan} but slightly  larger compared to the previous
MC estimate $\sf_{+-}(0) = 4.900(64)$~\cite{krech} 
(indicated as a full circle in
Fig.~\ref{fig:delta_fit}(b), still affected by
finite-size corrections)
and the analytical estimates $\sf_{+-}^\mathrm{(FT)}(0) = 3.16\ldots 4.78$. The
latter depend on the approximant used to re-sum the 
$O(\epsilon)$ series (see Ref.~\cite{krech} for details).

By using the values of $\tilde g_1$ and $\tilde g_2$
 obtained  previously in the context of the
Casimir amplitude we determine the coefficient $g_{\omega}$ of the correction
to the scaling variable $x$ (see Eq.~\reff{eq:xy_x_fit} with $r_1=0$) 
in order to achieve a good  data collapse for the whole scaling function, with
the results $g_{\omega} = 2.04(15)$ for $(++)$ BC and 
$g_{\omega} = 2.90 (15)$ for $(+-)$ BC.
The comparison between three phenomenological ans\"atze for the corrections
to scaling, i.e., cases (i) [Eq.~\reff{eq:xy_f_fit1}], (ii)
[Eq.~\reff{eq:xy_f_fit2}], and  (iv) [Eq.~\reff{eq:fit_new}], 
are presented in
Figs.~\ref{fig:ising_pp} and~\ref{fig:ising_pm}  for
$(++)$ and $(+-)$ BC, respectively. 
The scaling functions corresponding to the rational
expression for the corrections to scaling ansatz (case (iv)) 
lie in between the two others.

Currently, for the film geometry with $(++)$ BC there are 
no experimental data available for comparison, but in Fig.~\ref{fig:ising_pp}
 $\sf_{++}$ can be compared
with the
prediction of mean-field theory~\cite{krech} (MFT, solid line, normalized
such that $\sf_{++}^\mathrm{(MFT)}(0) = \sf_{++}^\mathrm{(MC)}(0)$ [$=2 \Delta^{\mathrm{(MC)}}_{++}$ see
Fig.~\ref{fig:delta_fit}(a)]) and  with the prediction of the
two-dimensional
Ising model~\cite{ES} (dashed line).
Recently, the de Gennes-Fisher local-functional
method has been extended to  study the
three-dimensional case with $(++)$ BC~\cite{upton,bu-08}.
In this latter (non-perturbative) 
approach one takes advantage of the knowledge of the values of
{\it bulk} critical exponents and amplitude ratios in order to fix completely
certain parameters of an effective model which is then used to calculate the
structural 
properties and the free energy 
of the system first in the presence of a single  wall and eventually in thin
films, giving access to the scaling function for $(++)$ BC.
The resulting scaling function (dash-dotted line in Fig.~\ref{fig:ising_pp}) 
is in very good agreement with the 
one (bottom set of data points in Fig.~\ref{fig:ising_pp}) determined
numerically via MC simulations 
by assuming corrections to scaling of the form given by
Eqs.~\reff{eq:xy_f_fit1} and~\reff{eq:xy_x_fit} 
with $r_{1,2} = 0$ and suitable values for the
fitting parameters $g_\omega$ and  $g_1$ (see above).
This agreement suggests that corrections to scaling are properly captured by
such ans\"atze even for $L\rightarrow \infty$. The prediction 
of the de Gennes-Fisher local-functional
method for the critical Casimir amplitudes (shown as diamonds in
Figs.~\ref{fig:delta_fit}(a) and (b)) is
$\Delta_{++} =  -0.42(8)$  and $\Delta_{+-} =  3.1$ \cite{upton},  which
compares  quite well with our MC
results for  $\Delta_{++} =  -0.376(29)$, whereas for  $\Delta_{+-}$
the agreement with our result   $\Delta_{+-} =  2.71(2)$  is slightly less good.
\begin{figure}
\includegraphics[height=7cm,width=0.45\textwidth]{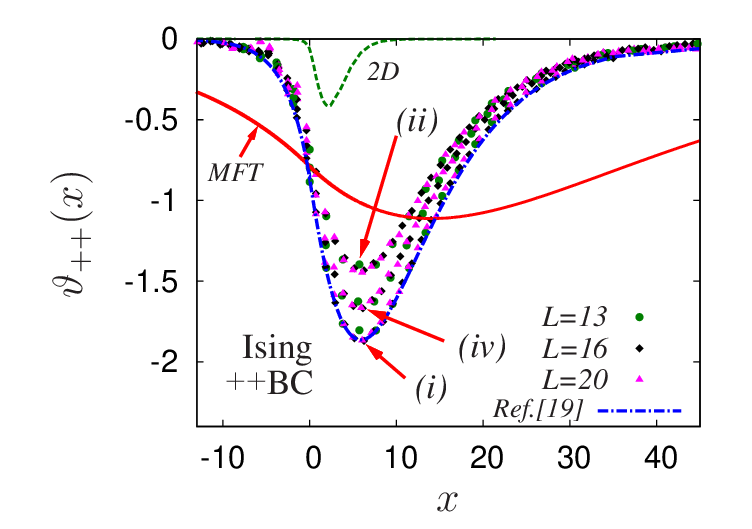}
\caption{%
Scaling function $\sf_{++}$ of the critical Casimir force in the three-dimensional Ising model with $(++)$ BC and zero bulk field. 
Data points refer to lattices with fixed inverse aspect ratio $1/\rho =6$.
The bottom and top data sets  have been obtained by accounting for corrections to
scaling according to Eq.~\reff{eq:xy_f_fit1} (case (i))
and Eq.~\reff{eq:xy_f_fit2} (case (ii)), respectively. 
The intermediate data set, instead, considers corrections of 
the rational form given by  Eq.~\reff{eq:fit_new} (case (iv)).
In each case the data collapse turns out to be very good within the range of the
scaling variable $x$ covered in the figure. 
The final estimate of the scaling function is biased by the functional form
assumed for the corrections to scaling. The position $x_\mathrm{min}\simeq 5.90(8)$ of the minimum is insensitive with respect to  these choices for the form of the corrections.
For comparison we provide the prediction of mean-field theory \cite{krech}
(solid line), normalized such that $\sf_{++}^{\mathrm{(MFT)}}(0)=\sf_{++}^{\mathrm{(MC)}}(0)$ [Fig.~\ref{fig:delta_fit}(a)],  the exact result for the
two-dimensional Ising model \cite{ES} (dashed line), and the result from the extended de Gennes-Fisher local-functional method \cite{bu-08} (dash-dotted line). Note that the actual phase transition of the film occurs at a  nonzero value of the bulk field. 
}
\label{fig:ising_pp}
\end{figure}

In the case of $(+-)$ BC we can compare
$\sf_{+-}$ with the experimental results
of Ref.~\cite{pershan}, with the prediction of
mean-field theory~\cite{krech} 
and with the corresponding result for the two-dimensional Ising
model~\cite{ES} (see Fig.~\ref{fig:ising_pm}). 
The  solid line, normalized similarly as for $(++)$ BC, represents 
the MF result, whereas the dashed line refers to
the two-dimensional Ising model~\cite{ES}. 
We expect the experimental data in Ref.~\cite{pershan} to
be affected by corrections to scaling already for $x\gtrsim 2$, 
due to the relatively small corresponding value of
$\xi/\ell \lesssim 30$, where $\ell \simeq 3$\AA\ is the molecular
scale set by the specific binary liquid mixture used in Ref.~\cite{pershan}.
In view of these difficulties, the comparison  between the MC and the
experimental data in Fig.~\ref{fig:ising_pm} can be regarded to provide
an encouraging agreement.
Within the Derjaguin approximation our numerical results for $\sf_{++}$ and
$\sf_{+-}$ form the basis for the calculation~\cite{Hertlein}  of the
corresponding scaling functions for the critical Casimir potentials in 
the sphere-plate
geometry, which turn out to be in
remarkably good agreement with the actual experimental results for that
geometrical setting~\cite{Hertlein}.

Comparing the scaling functions for $d=2$, $3$, and $4$ (MFT) one finds
that in the case of $(++)$ [$(+-)$] BC the position of the minimum [maximum]
moves away from the bulk critical point $x=0$ as the spatial dimension
increases. For $(++)$ [$(+-)$] BC the minimum [maximum] occurs above [below]
$T_c$ for all $d$. The {\it shapes} of the scaling functions in $d=2$ and
$d=3$ exhibit an interesting resemblance. 

As we pointed out above, in the case of $(+-)$ BC  
the fluctuations of the order parameter
are enhanced by the presence of a strongly
fluctuating interface in the middle of the film. This results in a critical 
Casimir force which is generally
stronger than in the case of $(++)$ BC, for which there is no such an
interface.
This is reflected by the fact that the amplitude $\sf_{+-}$ is larger than that
of $|\sf_{++}|$, e.g.,
$\sf_{+-}^{\mathrm{(max)}}/|\sf_{++}^{\mathrm{(min)}}| \simeq 3.8$ for the
data sets obtained by accounting for the corrections to scaling according to 
 Eq.~\reff{eq:xy_f_fit1} (case (i))
and Eq.~\reff{eq:xy_f_fit2} (case (ii)).
Even though field-theoretical 
MFT {\it per se} does not provide  quantitative predictions for
the overall amplitudes of the scaling functions $\sf_{++,+-}$, it yields
the relation~\cite{footnote2} 
\be
\sf_{+-}^{\mathrm{(MFT)}}(x) = - 4 \sf_{++}^{\mathrm{(MFT)}}(-2
x)
\ee
and therefore predicts 
$\sf_{+-}^{\mathrm{(max)}}/|\sf_{++}^{\mathrm{(min)}}| = 4$ and that the
maximum of $\sf_{+-}$ [minimum of $\sf_{++}$] occurs below [above] $T_c$.
Thus MFT captures already quite well the qualitative and quantitative
differences due to the presence or absence of an interface in the film.
In addition, the fluctuations of such an interface, occurring in particular 
at low temperatures, cause the scaling function $\sf_{+-}$ to decay to zero for
$x\rightarrow -\infty$ more slowly than the scaling function $\sf_{++}$, which
is clearly visible 
by comparing Figs.~\ref{fig:ising_pp} and~\ref{fig:ising_pm}.
%
\begin{figure}
\includegraphics[height=7cm,width=0.45\textwidth]{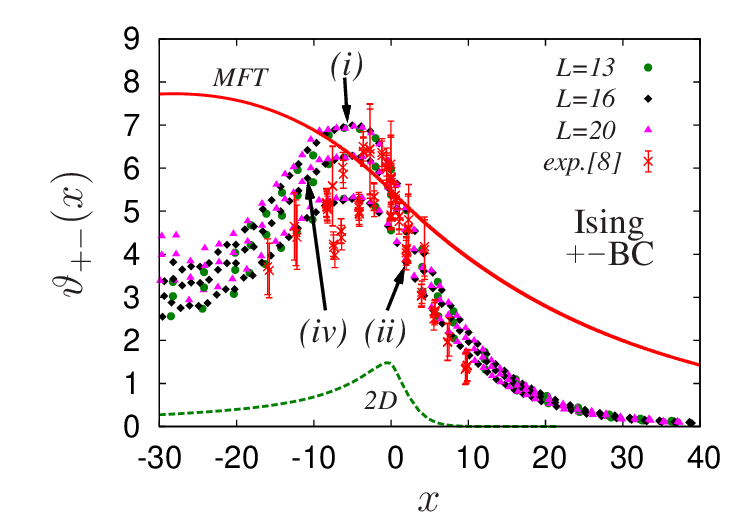}
\caption{%
Scaling function $\sf_{+-}$ 
of the critical Casimir force in the three-dimensional Ising model
 with $(+-)$ BC  and zero bulk field. Data points refer to lattices with fixed inverse aspect ratio $1/\rho =6 $.
For comparison we provide the mean-field
prediction \cite{krech} (solid line), normalized such that
 $\sf_{+-}^{\mathrm{(MFT)}}(0)=\sf_{+-}^{\mathrm{(MC)}}(0)$ [$=2\Delta_{+-}$, Fig.~\ref{fig:delta_fit}(b)], 
the exact result for the
two-dimensional Ising model \cite{ES} (dashed line), and the set of experimental data points
from Ref.~\protect{\cite{pershan}}.
The top and bottom data sets have been obtained by accounting for corrections to
scaling according to Eq.~\reff{eq:xy_f_fit1} (case (i))
and  Eq.~\reff{eq:xy_f_fit2} (case (ii)), respectively. 
The intermediate data set, instead, considers corrections of 
the rational form given by Eq.~\reff{eq:fit_new} (case (iv)).
In each case the data collapse turns out to be very good for $x\ge -20$. 
The final estimate of the scaling function is biased by the functional form
assumed for the corrections to scaling. The position $x_\mathrm{max}\simeq -5.4(1)$ of the 
maximum  is insensitive with respect to  these choices for the form of the corrections.
In spite of this caveat the comparison with the experimental data is
encouraging. Note that the actual phase transition of the film occurs at a nonzero value of the bulk field.
}
\label{fig:ising_pm}
\end{figure}

\subsubsection{Dirichlet-Dirichlet boundary conditions}
\label{subsec:DD}

\begin{figure}
\includegraphics[height=7cm,width=0.45\textwidth]{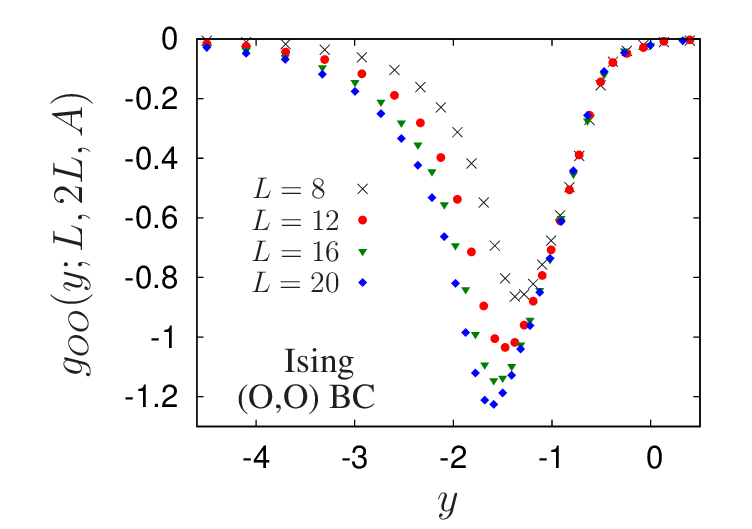}
\caption{Monte Carlo data for
 $g_{OO}(y=\tau(L-\frac{1}{2})^{1/\nu};L,2L,A=(L/\rho)^2)$ (see
 Eq.~\reff{eq:defg}, $\tau=(T-T_c)/T_c$) in the three-dimensional Ising model
 with $(O,O)$ BC for $L=8$, $12$, $16$, $20$, and for a fixed aspect ratio
 $\rho=1/6$.
 The 2D critical point of the film is located at $y=y_{c,OO}=-2.5(5)$, 
 as inferred from extrapolating the data in Table II of Ref.~\cite{KOI-96} to
 $L\rightarrow\infty$.
}
\label{fig:is_oo_gy}
\end{figure}

\begin{figure}
\includegraphics[height=7cm,width=0.45\textwidth]{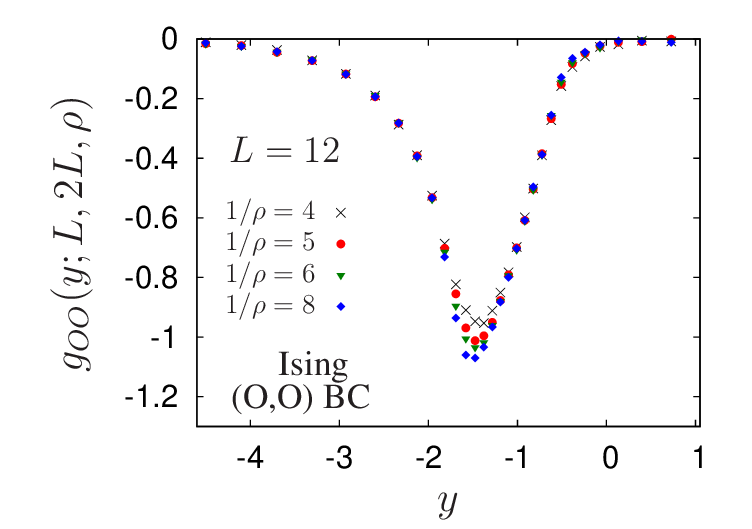}
\caption{Monte Carlo data for
  $g_{OO}(y=\tau(L-\frac{1}{2})^{1/\nu};L,2L,A=(L/\rho)^2)$ (see
  Eq.~\reff{eq:defg}, $\tau=(T-T_c)/T_c$) in the three-dimensional Ising model
  with $(O,O)$  BC for $L=12$ and various values of the  inverse aspect ratio
  $1/\rho={\sqrt A}/L$.
}
\label{fig:is_a}
\end{figure}

\begin{figure}
\includegraphics[height=7cm,width=0.45\textwidth]{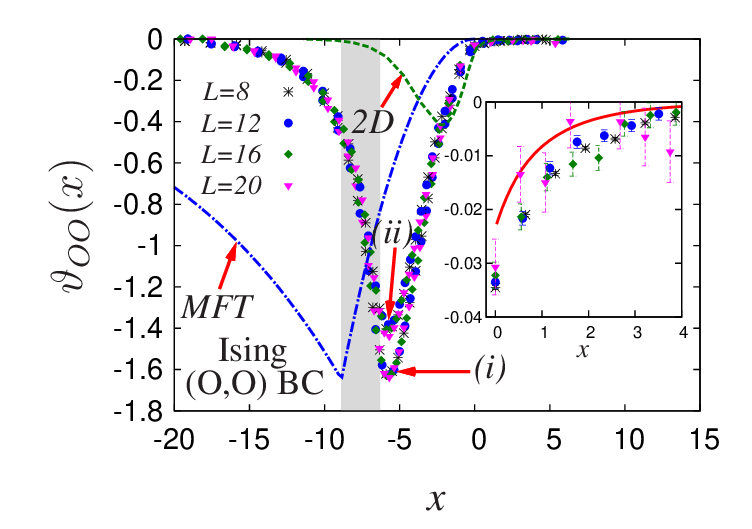}
\caption{Scaling function $\sf_{OO}$ of the Casimir force for the
three-dimensional Ising model with $(O,O)$ BC and zero bulk field.
The MC data refer to lattices with
$L=8$, $12$, $16$, $20$ and with a fixed inverse aspect ratio $1/\rho=6$.
Corrections to scaling have been accounted for according to two different
ans{\" a}tze, provided by Eq.~\reff{eq:xy_f_fit1} and 
 Eq.~\reff{eq:xy_f_fit2},  and
the corresponding numerical results are denoted by (i) and (ii), 
respectively.
With corrections of the form (ii), the {\it shape} of the resulting scaling
 function is almost indistinguishable from the one obtained with corrections
 of the form (i), but its overall amplitude  is reduced by a factor $R\simeq
 0.866$. For comparison we show the exact result
 for the  2D Ising model \cite{ES} (dashed line) and  the mean-field
prediction \cite{LGW-MF} (dash-dotted line) normalized such  that it yields
the same depth  of the minimum as the one
of the MC data (i). 
In the  inset we compare the MC data corresponding to the case (i)
with the scaling function 
obtained from the $\epsilon$-expansion \cite{KD-92}.
The gray bar indicates the value $x^*_{OO} = -7.6(1.3)$ (and its uncertainty) 
of the scaling
variable $x$ corresponding to the occurrence of the shifted critical point,
inferred from extrapolating the data in Table II of Ref.~\cite{KOI-96} to
$L\rightarrow\infty$.}
\label{fig:is_oo_scaling}
\end{figure}

In Fig.~\ref{fig:is_oo_gy}  we show the MC data  corresponding to 
$g(y;L,2L,A)$ (see Eq.~\reff{eq:defg}) for the Ising model with  $(O,O)$  BC,
realized by free surface spins.
The $L$-dependence of these data is quite pronounced and resembles that  
for the XY model with the same BC (compare Fig.~\ref{fig:xy_gy}(a)).
On the other hand,
the aspect ratio dependence appears to be relevant
only in the narrow interval $-2\lesssim y \lesssim -1$ 
(see Fig.~\ref{fig:is_a}), which is
similar to the case of the XY model with periodic BC 
(compare Fig.~\ref{fig:xy_a}).
As anticipated in Subsec.~\ref{subsec:XY}, the Ising model in a 3D film with 
Dirichlet-Dirichlet or periodic BC displays its 2D critical
behavior at a critical point which is located
on the bulk coexistence line $H=0$ at a size-dependent temperature
$T_c(L)$ such that $T_c(L\rightarrow\infty) = T_c (1 + y_c L^{-1/\nu})$
\cite{FSS}, where $y_c$ is a non-universal constant which depends, inter alia,
on the BC. From extrapolating the MC data for $T_c(L)$ reported in
Table II of Ref.~\cite{KOI-96} to $L\rightarrow\infty$ one infers
$y_{c,OO}=-2.5(5)$ for the Ising model with $(O,O)$ BC. 
As in the case of the XY model, the  residual dependence on $\rho$ observed in
Fig.~\ref{fig:is_a} might be due to the influence of the 2D phase transition
for $y\simeq y_{c,OO}$. Such a dependence cannot be captured by
ans\"atze such as the ones considered so far, which assume that 
the  corrections to scaling due to $\rho\neq 0$ are independent of $x$. 
Therefore, in order to achieve a good  collapse
of the data sets  corresponding to  different lattice sizes we  account
for corrections to  scaling by  following the procedure
applied to  the XY model with $(O,O)$ BC,  but we do not consider
an  aspect ratio dependence,
i.e., we use the ans\"atze  in Eqs.~\reff{eq:xy_x_fit},
~\reff{eq:xy_f_fit1} (case (i)), and ~\reff{eq:xy_f_fit2}   (case (ii)) with $r_{1,2}=0$.
As a result of the fitting procedure in the interval 
$x\in [-7,-4]$ we find 
$g_{1}=6.55(8)$  and $g_{\omega}=2.35(3)$ in case (i), and
$g_{2}=-2.877(15)$  and
$g_{\omega}=2.35(3)$ in case (ii). 
Figure~\ref{fig:is_oo_scaling} shows  the corresponding 
resulting estimates of the scaling function  $\sf(x)$ of the critical 
Casimir force with an excellent data collapse.  
As before, we find that  $\sf(x)$ is   affected by the choice of the functional
form of corrections to scaling. In the two cases (i) and (ii) 
one finds estimates
of $\sf(x)$ which have the same shape but the overall amplitude  is reduced by
a factor $R\simeq 0.866$ in case (ii) compared with case (i).
%

Due to the  residual dependence on the aspect ratio $\rho$,
$x_\mathrm{min}(\rho)$ and 
$\sf_\mathrm{min}(\rho)$ decrease upon decreasing $\rho$ and therefore the
values of $\sf_\mathrm{min}$ and $x_\mathrm{min}$ quoted above overestimate
the actual $\sf_\mathrm{min}(\rho = 0)$ and $\sf_\mathrm{min}(\rho = 0)$. 
The accuracy of our data does not allow us to study in more detail the Casimir
amplitude $\Delta_{O,O}\equiv \vartheta(0)/2$ (as we did
for $\Delta_{++}$ and $\Delta_{+-}$ in Fig.~\ref{fig:delta_fit}),
which turns out to be very small for $(O,O)$ BC. 
Indeed the estimate from the partially resummed 
$\epsilon$-expansion is
$\Delta_{O,O} = -0.0164$ \cite{krech}, 
whereas MC simulations yield $\Delta_{O,O} = -0.0114(20)$ \cite{krech}.
However, from our data for the scaling function 
we can estimate $\Delta_{O,O} = -0.014(8)$.
The corrections of form (i) yield for the 
 pronounced minimum of the scaling function 
$x^\mathrm{(i)}_\mathrm{min} = -5.74(2)$
and $\sf^\mathrm{(i)}_\mathrm{min}\equiv \sf(x^\mathrm{(i)}_\mathrm{min}) = -1.629(3)$
whereas those of form (ii) result in $x^\mathrm{(ii)}_\mathrm{min} = -5.73(4)$
and $\sf^\mathrm{(ii)}_\mathrm{min} \equiv \sf(x^\mathrm{(ii)}_{\mathrm{min}}) = -1.41(1)$. 
%
%

In 2D the scaling functions obey the 
relation $\sf_{OO}(x)=\sf_{++}(-x)$ \cite{ES}. 
We note that in 3D this relation  holds approximately for the positions
 of the minima of the scaling functions ($x^{(O,O)}_\mathrm{min}\simeq -5.7, x^{(+,+)}_\mathrm{min}\simeq 5.90$) but for the 
$(O,O)$ BC the scaling function vanishes more rapidly 
 than the scaling function
for the $(++)$ BC. For comparison in Fig.~\ref{fig:is_oo_scaling} 
 we provide the exact result for the two-dimensional
Ising model (dashed line).
In the inset we show our MC data corresponding
to the case (i) together with  the scaling function obtained by using
the $\epsilon$-expansion in Ref.~\cite{KD-92}. 
We note that it yields 
$\sf(0)/2 = \Delta_{O,O}\simeq - 0.0118$, which is larger than 
the estimate $\Delta_{O,O}\simeq -0.015$ given in
the same paper \cite{KD-92}, and obtained 
from dimensional interpolation; the latter value is still
larger than the
more recent theoretical 
estimate $\Delta_{O,O}\simeq -0.0164$ in Ref.~\cite{krech}.

In the case of Dirichlet-Dirichlet boundary conditions discussed here, the
film exhibits the 2D critical behavior at $T=T_c(L)$, corresponding to a
universal value $x^* = y_c (\xi_0^+)^{-1/\nu}$ of the scaling variable $x$. 
Close to the temperature $T_c(L)$, the free energy of the film is expected
to exhibit the singularity $\sim |T-T_c(L)|^{2-\alpha_{\mathrm{2D}}}$, where
$\alpha_{\mathrm{2D}}$ is the critical exponent of the specific heat of the
two-dimensional system.
This implies \cite{KD-92} that the scaling function $\vartheta_{OO}(x)$ of the
Casimir force displays a singularity  $\sim |x-x^*|^{2-\alpha_\mathrm{2D}}$ at
$x=x^*$, i.e., $\sim (x-x^*)^2 \ln |x-x^*|$ for the Ising model. This
singularity is too weak to be detectable by the present MC data.
In Fig.~\ref{fig:is_oo_scaling} the gray bar indicates the value of $x^*_{OO}=
y_{c,OO}(\xi_0^+)^{-1/\nu}= -7.6(1.3)$ and the
associated uncertainty. Accordingly, the singularity is expected to
occur on the left side of the pronounced dip.

So far there are no experimental data available that would correspond 
to the Ising universality class  with  $(O,O)$ BC. For experiments with 
binary liquid mixtures
the $(O,O)$ BC would  correspond to  walls which  have 
no adsorption preferences, i.e.,
both components of the mixture are attracted equally by each surface.
Effectively, in the limit of large film thicknesses, this can be achieved by
chemically decorating the confining walls by stripes of equal width and 
alternating preferences for the two species of the binary liquid mixture
(see Fig.~6 in Ref.~\cite{sprenger} for $S=1$, for which within MFT the effective
Casimir amplitude vanishes, corresponding to vanishing surface fields within
MFT so that $(O,O)$ BC hold).

%
%
%

\subsubsection{Periodic boundary conditions}
\label{subsec:IsP}

In the case of periodic BC the aspect ratio dependence of the Monte Carlo data
for the Ising model (as for the XY model discussed in Subsec.~\ref{subsec:XY})
turns out to be relevant only in the vicinity of the minimum of  
the function $g_P(y;L,2L,A)$ which is  associated with the 
finite-size effects close to the actual critical point of the thin film.
Extrapolating the data in Table I of Ref.~\cite{KOI-96} to
$L\rightarrow\infty$ one infers that the shifted critical point corresponds to
$y=y_{c,P}=-1.60(2)$.
The fact that this type of finite-size dependence  
does not occur for the Ising model with fixed BC 
(see Fig.~\ref{fig:ising_asp_pp})
might be related to the different phase behavior below $T_c$ in 
the latter  case. 
For $(++)$ BC the critical point is shifted off the 
bulk coexistence line $H=0$ to some value
$(T_c(L),H_c(L))$  \cite{evans} and hence in the vicinity of the
minimum of the function $g_{++}(y;L,2L,A)$ the 
corresponding bulk correlation length
is smaller than the characteristic transverse length $L_\|=\sqrt A$. 
As already mentioned earlier, for  $(+-)$ BC below $T_c$ (but above the
temperature of unbinding of this interface from one or the other surface) 
there exists a single  film phase characterized by the  OP profile displaying
an interfacelike structure centered at the middle of the 
film~\cite{parryevans,binder}. In this film phase  
the parallel correlation function $\xi_{\parallel}$
governing the exponential decay of correlations along the interface is very
large even for temperatures further away from $T_c$, i.e., $\xi_{\parallel}
\sim \exp (L_{\parallel}/(4\xi))$ with $L_{\parallel} = L_x =
L_y$. $\xi_{\parallel}$ gives rise to the aspect ratio dependence of the
function $g_P(y;L,2L,A)$ for $y\lesssim -4$.

\begin{figure}
\includegraphics[height=7cm,width=0.45\textwidth]{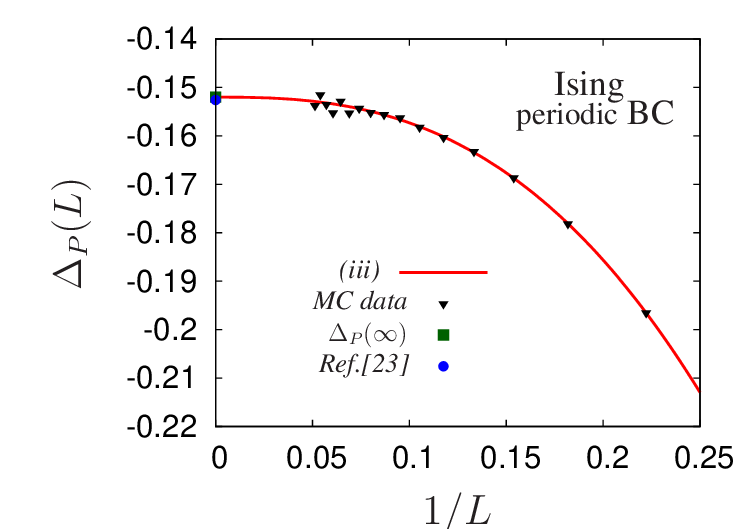}
\caption{%
MC data for the critical Casimir amplitude 
$\Delta_{\mathrm{P}}(L)$ for the three-dimensional Ising model with
periodic BC, as a function of the inverse film thickness $L$ (on lattices with
fixed inverse aspect ratio $1/\rho = 6$).
Due to $L$-dependent corrections to scaling, $\Delta_P$ depends on $L$
and reaches its asymptotic value  in the limit
$L\rightarrow\infty$. 
The solid line corresponds to the best fit obtained by using the fitting
ansatz given in Eq.~\reff{eq:xy_f_fit3} in the interval $0<1/L\le 0.25$.
Our estimate  ($\blacksquare$) for the asymptotic value of the Casimir amplitude
$\Delta_P(\infty)$ compares very well with the previous 
MC result ($\bullet$) from
Ref.~\protect{\cite{krech}}. 
}
\label{fig:ising_pbc_delta}
\end{figure}
%
In order to account for the corrections to scaling we 
follow the same procedure which 
we used for the XY model with periodic BC, i.e., 
we assume their $L$-dependence to be captured by Eq.~\reff{eq:xy_f_fit3} at
least within the range of sizes we are interested in.  
Accordingly, we focus on the data for the critical Casimir
amplitude $\Delta_P$ and we fit them according to Eq.~\reff{eq:fit_delta_new}
(case (iii), Eq.~\reff{eq:xy_f_fit3}).
The best fit parameters, based on all data points, are given 
by $g_3 = 16.10(55)$ and $\omegaeff = 2.664(27)$ and the resulting curve is
provided as a solid line in Fig.~\ref{fig:ising_pbc_delta}.
The associated estimate for the asymptotic value 
$\Delta_P(L\to\infty) \equiv \Delta_P= -0.1520(2)$  agrees very  well
with the MC result $-0.1526(10)$ from Refs.~\cite{krech,KL}.

\begin{figure}
\includegraphics[height=7cm,width=0.45\textwidth]{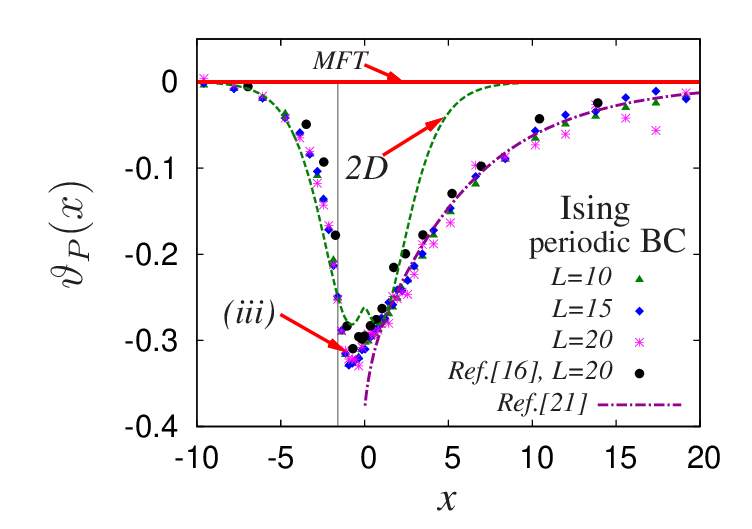}
\caption{%
Scaling function $\sf_{\mathrm{P}}(x)$ 
of the critical Casimir force in the three-dimensional Ising model
with periodic BC and zero bulk field. The data points refer to lattices with the inverse
 aspect ratio $1/\rho = 6$. The corrections to scaling are taken into account
according to Eqs.~\reff{eq:xy_f_fit} and \reff{eq:xy_f_fit3} (case (iii)), see the main text).
For comparison we show also the data
set corresponding to the lattice with thickness $L=20$ as investigated in Ref.~\protect{\cite{DK}},
the analytical prediction of Ref.~\cite{GD} (dash-dotted line) 
for $x\ge 0$, and results for 2D Ising model (dashed line) that we have
obtained numerically by using the transfer matrix method. 
Due to the self-duality of the 2D Ising model one has
  $\sf_{\mathrm{P}}(-x) = \sf_{\mathrm{P}}(x)$ for $d=2$ which allows
for the occurrence of  two symmetric minima \cite{rudnick}. We note that MFT
yields  $\sf_{\mathrm{P}}(x)\equiv 0$ (solid line).  
The gray vertical line indicates the universal 
value $x^*_P = -1.60(2)$ of the scaling variable $x$
corresponding to the occurrence of the shifted critical point,
inferred from extrapolating the data in Table I of Ref.~\cite{KOI-96} to
$L\rightarrow\infty$.
}
\label{fig:ising_pbc_scaling}
\end{figure}

The scaling function  $\sf_\mathrm{P}$ can now be determined by assuming that
Eq.~\reff{eq:xy_f_fit}, with $r_2=0$ and the parameters $g_3$ and
$\omegaeff$ obtained from the analysis of $\Delta_P(L)$, effectively describes 
its corrections to scaling (case (iii), Eq.~\reff{eq:xy_f_fit3}), which
actually leads to a very good data collapse in a wide  range of
temperatures.
It also turns out that no
corrections to the scaling variable $x$ (see Eq.~\reff{eq:xy_x_fit}) 
are required in order to achieve it, i.e., $r_1$, $g_\omega \simeq 0$.
(Note, however, that corrections due to $\rho\neq 0$ might be particularly
relevant within a certain range of the scaling variable $x$, see below.)
The resulting scaling function  
$\sf_{\mathrm{P}}$ is presented 
in Fig.~\ref{fig:ising_pbc_scaling} and it is based on a larger set of
geometries of the simulation cell and with a better accuracy than 
in our earlier work~\cite{EPL}. 
The scaling function is in very good agreement 
with its previous determination in Ref.~\cite{DK} 
based on the computation of the lattice stress tensor.
The slight discrepancies might be due to the uncertainty in the 
normalization factor which had to be used in Ref.~\cite{DK} (see also
Subsec.~\ref{subsec:XY}). 
This agreement provides additional support concerning the reliability
of our approach.

%

\begin{figure}
\includegraphics[height=7cm,width=0.45\textwidth]{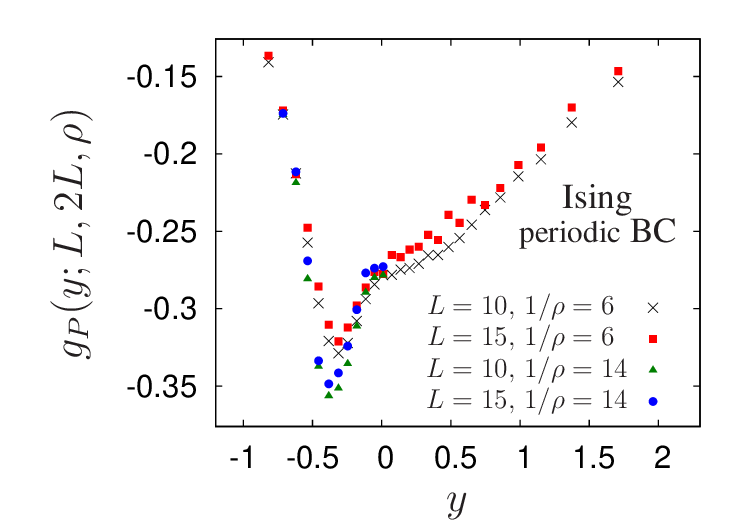}
\caption{%
Aspect-ratio dependence of the  function  $g_{\mathrm
  {P}}(y=\tau(L-\frac{1}{2})^{1/\nu};L,2L,A =(L/\rho)^2)$ 
(see Eq.~\reff{eq:defg}) for the three-dimensional Ising  model with  periodic
BC.
For a fixed value of $\rho$, this function depends only weakly on 
 $L$. 
By changing $\rho$, the function  $g_{\mathrm {P}}$
is affected mainly in  the region $-0.6\lesssim y \lesssim 0$. 
The 2D critical point of the film is located at $y=y_{c,P}=-0.52(2)$, 
as inferred from extrapolating the data in Table I of Ref.~\cite{KOI-96} to
$L\rightarrow\infty$.
}
\label{fig:ising_asp_per}
\end{figure}

Figure~\ref{fig:ising_pbc_scaling} presents also the comparison with the
analytical prediction of the recently proposed field-theoretical (FT) 
expansion up to $O(\epsilon^{3/2})$~\cite{GD} 
(dash-dotted line) for $x\ge 0$. 
This latter prediction is now in better agreement with the MC data than 
the previous
$O(\epsilon)$ field-theoretical result in Ref.~\cite{KD-92} but still misses
the onset of the formation of the minimum.
(Figure 5 in Ref.~\cite{EPL} compares the MC data with the $O(\epsilon)$
results, revealing a significant discrepancy
for $0 < x \lesssim 4$.)
The estimated value of 
$\vartheta^{\mathrm{(FT)}}_{P}(0) = -0.39$ from Refs.~\cite{DGS,GD} does not
agree with our  MC estimate $\vartheta_{P}(0)= -0.3040(4)$. 
For the minimum of the scaling function we find the estimates  
$x_\mathrm{min} = -0.681(1)$,
$\sf_\mathrm{min} \equiv \sf_\mathrm{P}(x_{\mathrm{min}}) = -0.329(1) $.
Note, however, that for $x \simeq x_\mathrm{min}$ the corrections due to
$\rho\neq 0$ are expected to be relevant. In order to substantiate this
statement we have determined the
 function  $g_{\mathrm {P}}(y;L,2L,A=(L/\rho)^2)$  
(see Eq.~\reff{eq:defg})  also 
from a set of data for lattices of
thickness $L=10$ and $15$ with an inverse 
aspect ratio $\rho^{-1} = 14$, which can be
compared with the corresponding data 
set from Fig.~\ref{fig:ising_pbc_scaling},
for which $\rho^{-1} = 6$. 
This comparison is presented in Fig.~\ref{fig:ising_asp_per} and 
clearly shows that, while the  function  $g_{\mathrm {P}}$ is actually only slightly
dependent  on  the
thickness  $L$ of the lattice, 
there is a  dependence on $\rho$ which, however,  is
relevant only very close to $x_\mathrm{min}$. As mentioned above for the case of $(O,O)$ BC,
 this latter dependence on $\rho$  cannot be captured by
ans\"atze such as the ones considered so far because they assume
$x$-independent corrections due to $\rho\neq 0$.
Therefore, similar to the case of $(O,O)$ BC, due to the residual dependence
on  $\rho$, $x_\mathrm{min}(\rho)$ and
$\sf_\mathrm{min}(\rho)$ decrease upon decreasing $\rho$ and therefore the
values of $\sf_\mathrm{min}$ and $x_\mathrm{min}$ quoted above overestimate
the actual values of  $\sf_\mathrm{min}(\rho = 0)$ and $x_\mathrm{min}$.

%
%
%
%
%
%
As in the case of Dirichlet-Dirichlet boundary conditions,  the
point at which the film exhibits the 2D critical behavior is located
on the bulk coexistence line and corresponds to a value $x^*$ of
the scaling variable $x$, at which the scaling function is expected to
display
the weak singularity $\sim (x-x^*)^2 \ln |x-x^*|$ (see Sec.~\ref{subsec:DD}
above). In Fig.~\ref{fig:ising_pbc_scaling}
the gray vertical line indicates the corresponding universal value
$x^*_P=y_{c,P}(\xi_0^+)^{-1/\nu} = -1.60(2)$. 
Accordingly, also in this case the singularity is expected to occur on the
left side of the pronounced dip but cannot be detected by the present MC
data.

%
%
%
%

\begin{figure}
\includegraphics[height=7cm,width=0.45\textwidth]{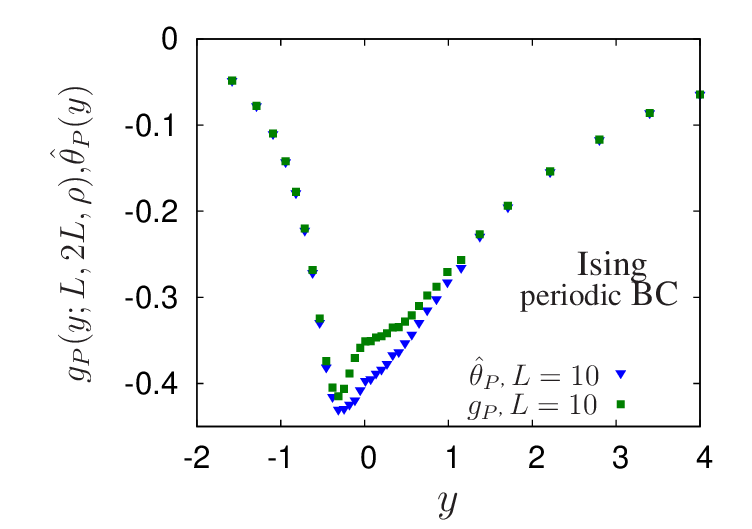}
\caption{%
Plot of the function $g_{\mathrm{P}}(y;L,2L,A=(L/\rho)^2)$  
(see Eq.~\reff{eq:defg}) and the associated scaling function 
$\sfhb(y)$ (i.e., the lattice
estimate of $\sfb(y) \equiv \sf_\mathrm{P}(y/(\xi_0^+)^{1/\nu})$, which is
calculated by solving  Eq.~\reff{eq:defg}) iteratively
for the three-dimensional 
Ising  model with  periodic  BC. The data points refer to a lattice with $L=10$
and $1/\rho=6$. The data have not been corrected for the 
corrections to scaling.
The pronounced shoulder originally present in $g_P$ is smoothed out upon
calculating the associated scaling function.
}
\label{fig:ising_pbc_restore}
\end{figure}

As a final remark we point out that for
the Ising model with periodic BC the function $g_P(y;L,2L,A)$
exhibits a somewhat peculiar   shape near $T_c$ with a
characteristic ``shoulder'' formed above the critical temperature.
The  procedure for retrieving  the  scaling function $\sfhb(y)$ [the lattice
estimate of $\sfb(y) \equiv \sf_\mathrm{P}(y/(\xi_0^+)^{1/\nu})$]
via Eq.~\reff{eq:implicit} involves rescaling of the argument 
of the scaling function and removes the ``shoulder'' structure
from the curve. The formation of this ``shoulder'' is related to
 the particular shape of $\sf_{\mathrm {P}}$ which 
on  the left side of  the minimum increases more steeply than 
on  the right side of it (see Fig.~\ref{fig:ising_pbc_restore}).

%
%
%
%
%
%
%

\section{Summary and conclusions}
\label{sec:concl}
\begin{description}
{\bf \item{ A.}} {\bf Summary}
\end{description}

We have presented important details of 
a novel general approach~\cite{EPL} to determine the universal 
scaling functions $\sf$ of critical Casimir forces via MC simulations. 
We have applied this method (see Subsects.~\ref{subsec:fe} and \ref{subsec:determ} as well as Fig.~\ref{fig:fig1})
in order to study the scaling
functions corresponding to the three-dimensional Ising and XY bulk
universality
classes for a variety of universal 
boundary conditions in film geometries with varying thickness
$L$.
Corrections to scaling appear to be quite relevant in
the range of sizes $L$ we have investigated, which are strongly limited by the
steeply increasing computational  costs required for larger systems. 
In spite of these difficulties, it is possible to analyze the corresponding MC
data by assuming suitable ans\"atze for corrections to scaling. Even if  the
final numerical determinations of the scaling functions are biased by these
assumptions,  they turn out to be consistent with the results of different
numerical and analytical approaches and with all available experimental data.

Our main results are the following:

(1) We have obtained the Casimir scaling function 
 $\sf_{OO}$ for the three-dimensional 
XY model with $(O,O)$ BC  [(Dirichlet, Dirichlet) BC] 
(Fig.~\ref{fig:xy_scaling}). 
Corrections to scaling have been
 accounted for by using  two different
ans\"atze, provided by Eq.~\reff{eq:xy_f_fit1} (case (i)) and  
Eq.~\reff{eq:xy_f_fit2} (case (ii)).
These choices of the functional form of corrections to scaling
have been  dictated by the pronounced  dependences on
 $L$ and on the aspect ratio $\rho$ of the simulation cell
which  occur  for this type of  BC 
(Fig.~\ref{fig:xy_gy}(a)). 
Both ans\"atze lead to  a very good data collapse but the overall amplitude
of the scaling function is reduced by a factor $R\simeq 0.9$ 
in case (ii) compared to case (i).
Our MC data compare very well with the corresponding experimental data
for $^4$He films from Ref.~\cite{garcia}  and with the MC data of Ref.~\cite{hucht}. For comparison also mean field results are  provided.

(2) The Casimir scaling function $\sf_{P}$  and  the critical 
Casimir amplitude $\Delta_P$  have been obtained
 for the three-dimensional XY model
 with periodic BC 
 (Figs.~\ref{fig:xy_scaling_per} and  \ref{fig:delta_fit_per}). 
In this case, judged by the behavior of the generating
function $g$ introduced in  Eq.~(\ref{eq:defg}), corrections to scaling are much less pronounced
 than in the 
case of $(O,O)$ BC (Fig.~\ref{fig:xy_gy}(b)) and the aspect ratio dependence 
is relevant only in the restricted range of the scaling 
variable near the minimum of the scaling function   (Fig.~\ref{fig:xy_a}).
A very good data collapse is achieved by using  the ansatz with the
 effective exponent $\omegaeff=2.59(4)$
 (Eq.~\reff{eq:xy_f_fit3} (case (iii)) and by neglecting the 
corrections to scaling due to the aspect ratio dependence
 ($r_{1,2}=0$ in Eqs.~\reff{eq:xy_x_fit}  and \reff{eq:xy_f_fit}). 
The {\it shape} of our MC data agree very well with the
corresponding  MC data of Ref.~\cite{DK} which, however, have left the
amplitude undetermined.  Our estimate for the critical  Casimir amplitude is
$\Delta_P = -0.2993(7)$.
By extending the line of arguments of Ref.~\cite{kardar:04} to the present
case, we have theoretically predicted the value 
$\sf_P^{\rm (TH)}(-\infty) = -\zeta(3)/\pi \simeq -0.38$ [see
  Eq.~\reff{eq:XYperplat}] at which the scaling function $\sf_P(x)$ saturates
for $x\rightarrow -\infty$. This value is confirmed by the corresponding
estimate $-0.383(4)$ based on our MC data.

(3) We have obtained the scaling functions $\sf_{++}$,  $\sf_{+-}$ 
and the  corresponding
 Casimir amplitudes $\Delta_{++}, \Delta_{+-}$ 
of the critical Casimir force in the three-dimensional Ising model
 with $(++)$ and $(+-)$ BC, respectively, applicable for classical fluids
 (Figs.~\ref{fig:ising_pp}, \ref{fig:ising_pm} and \ref{fig:delta_fit}).
We find that in the  critical regime 
the numerical data are practically independent 
of the aspect ratio $\rho$ (Fig.~\ref{fig:ising_asp_pp}) but $L$-dependent corrections to scaling are rather important
(Fig.~\ref{fig:delta_fit}). 
The presented scaling functions and Casimir amplitudes
  have been obtained by accounting
 for corrections to
scaling according to Eq.~\reff{eq:xy_f_fit1} (case (i)), 
Eq.~\reff{eq:xy_f_fit2} (case (ii)), and  Eq.~\reff{eq:fit_new} (case (iv))
 with $r_{1,2}=0$ (thus neglecting the dependence of the data on $\rho$).
The final estimate of the scaling function is biased by the functional form
assumed for the corrections to scaling; all considered cases provide
 a very good data collapse.
The fitting
ansatz in Eq.~\reff{eq:fit_delta_new_r} describes very well the data for  the 
Casimir amplitudes $\Delta_{++/+-}$ as a function of the film thickness $L$.
Our estimates  for the asymptotic values of the Casimir amplitudes are
$\Delta_{++} = -0.376(29)$ and $\Delta_{+-} = 2.71(2)$ which 
compare reasonably well with previous MC results  from
Ref.~\protect{\cite{krech}} and with the results from the de Gennes-Fisher
local-functional approach \cite{upton}. 
Our results for the case of $(+-)$ BC compare well with 
recent X-ray scattering data for critical films of a classical 
binary liquid mixture \protect{\cite{pershan}}.
Moreover the MC data for the scaling functions
$\sf_{++}$ and  $\sf_{+-}$ have been  used
to calculate, within the Derjaguin approximation,
the corresponding scaling functions for the critical Casimir potentials
for the experimentally relevant geometry of a sphere near a
planar substrate. These  numerical results agree remarkably well 
with the experimental data for colloidal particles immersed 
in a critical solvent and close to a container wall \cite{Hertlein}.

(4) We have obtained the Casimir scaling function $\sf_{OO}$  for the
three-dimensional Ising model with 
 $(O,O)$ BC (Fig.~\ref{fig:is_oo_scaling}).
For these BC the $L$-dependence of the MC simulation data is quite
pronounced  (Fig.~\ref{fig:is_oo_gy}), 
similarly to the case of the XY model with the same $(O,O)$ BC.
The  dependence on the  aspect ratio is relevant only in the small range of the scaling variable near the minimum of the scaling function 
 (Fig.~\ref{fig:is_a}). Our data do not allow us to obtain a
quantitatively accurate estimate of the 
Casimir amplitude, because of its  very small value. 
 Corrections to scaling have been accounted for according to the
ans{\" a}tze provided by Eq.~\reff{eq:xy_f_fit1}  (case (i)) and 
Eq.~\reff{eq:xy_f_fit2} (case (ii)) 
 with $r_{1,2}=0$ (thus neglecting the  dependence of the data on the aspect
ratio $\rho$).

(5) The scaling function $\sf_{\mathrm{P}}(x)$ 
and the critical Casimir amplitude 
$\Delta_{\mathrm{P}}$ have been obtained for the three-dimensional Ising model with
periodic BC  (Figs.~\ref{fig:ising_pbc_scaling} and \ref{fig:ising_pbc_delta}).
As in the case of the XY model with  periodic BC, the aspect 
ratio dependence of the MC data appears to be pronounced 
only near the actual critical point 
of the thin film (Fig.~\ref{fig:ising_asp_per}). Therefore, the corrections to scaling have been accounted for in the same way as for the XY model with periodic BC, i.e., according to Eqs.~\reff{eq:xy_f_fit} and \reff{eq:xy_f_fit3} (case (iii)).
The best fit for the $L$-dependence of the Casimir amplitude
$\Delta_{\mathrm{P}}$ has been obtained by using the 
ansatz given in Eq.~\reff{eq:fit_delta_new_r}.
Our improved estimate for the  value of the Casimir amplitude
$\Delta_P = -0.1520(2)$ agrees very well with the previous MC result  from
Refs.~\cite{krech,KL}. The particular shape of the scaling 
function $\sf_{\mathrm{P}}(x)$ around its minimum is reflected in the formation
of a characteristic  ``shoulder'' in 
the corresponding generating function $g_P$  above the critical temperature (Fig.~\ref{fig:ising_pbc_restore}).

\begin{description}
{\bf \item{ B.}} {\bf Conclusions and outlook}
\end{description}

Our approach can be applied in order to study other experimentally relevant
geometrical settings as well as the effect of chemically or geometrically 
inhomogeneous confining surfaces on the critical Casimir force. 
In the latter cases, even lateral
critical Casimir forces are expected to act 
in addition to the normal Casimir force investigated here.
This lateral force has been theoretically investigated for 
chemically~\cite{SSD-06} and topographically~\cite{THD-08} patterned
surfaces, whereas it has been experimentally studied
for colloidal particles exposed to 
chemically patterned surfaces~\cite{CPS}. 

In addition to appliying our quantitative method to these cases, 
it is also desirable to perform more extensive and larger scale  MC
simulations in order to identify the origins of the corrections to scaling
and to characterize them more accurately, possibly to the extent which is by
now achieved for bulk critical phenomena.
This valuable knowledge would therefore allow an unbiased and thus even more
accurate
determination of the scaling functions of the critical 
Casimir force beyond the results
presented here.
Finally, beyond the application to thin $^4$He films near the
superfluid-normal fluid transition, our results for the three-dimensional XY
model with $(O,O)$ BC could be relevant for critical Casimir forces acting on
Bose-Einstein condensates~\cite{martin}.

%

\acknowledgments
The authors acknowledge the important contribution of M. De Prato
to the early stages of this work.
They are grateful to A. Hucht, M. Krech, and E. Vicari for useful
discussions and to  M. Fukuto and R. Garcia for providing their
experimental data.
In the context of the KITP program on the theory and practice
of fluctuation-induced interactions at the University of California, Santa Barbara, AG, AM, and SD were supported in part by the US National Science Foundation under Grant No. NSF PHY05-51164. 

%

\begin{appendix}   
\section{Corrections to scaling and  fitting procedures.}
\label{sec:size_corr} 

In this appendix we describe the general strategy we have used
in order to obtain the best fitted values of the parameters which control
the corrections to scaling. 
The main problems one faces are to quantify the quality of a certain data
collapse
and then to choose the parameters which influence it in such a way as 
to optimize this quality.
The estimation of the parameters and of the associated confidence
interval proceeds as in the case of least-square fits with chi-square tests
of the quality of the 
fit, but with the additional complication that the fitting
function itself is not known and has to be estimated from the numerical data
itself.

In what follows we describe the procedure we have used in order to determine
the best fit parameters which control the $L$-dependent corrections to
scaling. On the
same footing we have also treated the corrections due to a nonzero aspect
ratio $\rho\neq 0$ (see
Subsec.~\ref{subsec:corr}).
In full generality, assume that one seeks to determine, e.g., via MC
simulations, the finite-size scaling function $h$ of a quantity $\psi$ which,
in the absence of corrections to scaling, is expected to be a function of a
scaling variable $x$ only (which involves a suitable combination of
temperature and size $L$ of the system) so that
\be
\psi (x,L\to\infty) = h(x).
\ee
For the time being we omit possible algebraic $L$-dependent
prefactors of $h$.
In the MC simulations one considers a set of $N$ lattices of sizes 
$L_1$, $L_2$, \ldots, $L_N$ 
and by varying the temperature one collects for each size
$L_k$ a discrete set of numerical values $\psi_{k,j}$ of $\psi$ with 
$j=1,2  \ldots, j_k^\mathrm{max}$ which correspond to values $x_{k,j}$ of the
scaling variable $x$ in the interval $[x_\mathrm{min},x_\mathrm{max}]$. 
In this process the statistical uncertainty 
$\Delta\psi_{k,j}$ associated with $\psi_{k,j}$ is also determined.
From these quantities $\psi_{k,j}$ one intends
to determine $h$, taking into account the
presence of corrections to scaling. Due to them, $\psi$ is actually not a
function of $x$ only, but also of the size $L$ of the system. 
In order to cope with this one
therefore assumes the following functional structure:
\be
\label{eq:apfit}
\psi(x;L)=f_1(L;t_1)h(f_2(L;t_2) x)
\ee
where $f_1(L;t_1)$ and $f_2(L;t_2)$ capture the effects of the correction to
scaling on the quantity $\psi$ itself and on the scaling variable $x$,
respectively. These functions depend on the size $L$ of the system and on
certain
parameters $t_1, t_2$ which one would like to determine in such a way as 
to achieve the
best data collapse for the function $h$, obtained from the set of
data points
$(f_2(L_k;t_2) x_{k,j}, [f_1(L_k,t_1)]^{-1}\psi_{k,j}) =:
(y_{k,j}(t_1,t_2),h_{k,j}(t_1,t_2))$
for the various values of $j$ and $k$, and as to take  also into account the
statistical error $\Delta h_{k,j}(t_1) :=
[f_1(L_k,t_1)]^{-1}\Delta\psi_{k,j} $ associated with $h_{k,j}(t_1,t_2)$.

For each value $L_k$ we have interpolated the data set 
$(x_{k,j},\psi_{k,j})$  in the interval $[x_\mathrm{min},x_\mathrm{max}]$ 
by using a cubic spline approximation. This way we have constructed a function
$\psi_k(x)$ with $x\in [x_\mathrm{min},x_\mathrm{max}]$ and with 
$\psi_k(x_{k,j}) = \psi_{k,j}$.
From this function we have calculated 
the corresponding $L_k$-dependent estimate
$h_k(x;t_1,t_2)$ of $h$, given by
\be
h_k(y;t_{1},t_{2}) = f_1(L_k;t_1)^{-1}
\psi_k(y f_2(L_k;t_{2})^{-1}),
\ee
which fulfills $h_k(y_{k,j}(t_1,t_2);t_1,t_2) = h_{k,j}(t_1,t_2)$.
In order to assess the quality of the data collapse and the quality of the
fit we have actually to specify the function with which we would like to fit
the data, which is the yet unknown scaling function $h$. 
In order to achieve this, we define an expected 
model function $h_{\mathrm{expect}}$ as the average of the various $h_k$,
\be
h_{\mathrm{expect}}(y;t_{1},t_{2})=\frac{1}{N}\sum \limits_{k=1}^{N}
h_{k}(y;t_{1},t_{2}),
\ee
which will then be fitted 
to the observed MC values by adjusting the parameters $t_1$ and $t_2$.

Accordingly, we calculate the ``$\chi^2(t_{1},t_{2})$'' associated with
the fitting of the data points $(y_{k,j}(t_1,t_2),h_{k,j}(t_1,t_2))$ with the
function $h_{\mathrm{expect}}(y;t_{1},t_{2})$:
\be
\begin{split}
& \chi^2(t_{1},t_{2}) = \\
&\sum \limits_{k=1}^{N} 
\sum \limits_{j=1}^{j_k^\mathrm{max}} 
\frac{[h_{k,j}(t_1,t_2) -
    h_\mathrm{expect}(y_{k,j}(t_1,t_2);t_{1},t_{2})]^2}{[\Delta
    h_{k,j}(t_1)]^2}.
\end{split}
\ee
Due to the non-trivial and non-linear dependence of the fitted data 
(and of the fitting function) on the
parameters $t_i$ we cannot assume this quantity to play the same role as a
$\chi^2$ in more standard fitting procedures in which only the fitting function
depends on the parameters one wants to estimate. 
Nevertheless, we have heuristically made this assumption, i.e.,  that $\chi^2(t_{1},t_{2})$ plays the same role as a $\chi^2$, in order to
determine the best fit parameters and the associated confidence intervals.
Accordingly we have proceeded as usual by determining the optimal 
fit parameters $\bar
t_1$ and $\bar t_2$ which minimize the value of $\chi^2$: $\chi^{2}(\bar
t_{1},\bar t_{2})=\min_{\{t_1,t_2\}} \chi^{2}(t_{1},t_{2})$. 
In order to estimate the statistical uncertainty  $\Delta \bar t_i$ of 
$\bar t_{i}$ we have determined that region of the plane $(t_{1},t_{2})$
for which $\chi^{2}(t_{1},t_{2})< \chi^{2}(\bar t_{1},\bar
t_{2})+2.3$~\cite{Bev}.
The projection of the resulting region (typically of the form of an ellipse)
onto the axis $t_i$ gives $2 \Delta \bar t_i$, so that the estimate for the
parameters is of the form $\bar t_i \pm \Delta \bar t_i$.

\end{appendix}

%

\vfill\eject


\begin{thebibliography}{99}
\bibitem{footnote1}
Earlier observations~\protect{\cite{bey}} of flocculation in colloidal
suspensions close to the critical demixing point of a binary liquid mixture
suggested the action of critical 
Casimir forces; however, an unquestionable conclusion could not be
reached.
\bibitem{footnote}
The experimental value 1.2 \AA\ used in Ref.~\cite{LGW-MF1}
is not the correct one for the comparison presented therein. The correct
comparison of the experimental data with the mean-field analytical prediction
\cite{LGW-MF,LGW-MF1} and with the results of Monte Carlo simulations is
presented here and in Refs.~\cite{LGW-MF,EPL,hucht}.
\bibitem{footnote2} 
This relation between $\sf_{++}^{\rm (MFT)}(x)$ and 
$\sf_{+-}^{\rm (MFT)}(x)$ has seemingly been
overlooked in the literature even though it follows after some algebra from the
analytic expressions of the scaling functions reported in
Ref.~\protect{\cite{krech}}. 
In particular it also follows that $\sf^{\rm (MFT)}_{++}$ has
a minimum for $x = x_\mathrm{min}^\mathrm{(MFT)} = 14.05$, whereas
$\sf^{\rm (MFT)}_{+-}$ has a maximum for 
$x = - 2  x_\mathrm{min}^\mathrm{(MFT)} \simeq
28.1$ (which improves the estimates reported in the caption of Fig.~1 in
Ref.~\protect{\cite{krech}}).
\end{thebibliography}
\end{document}